\documentclass[11pt]{article}
\usepackage{amssymb}

\usepackage[totalwidth=460truept,totalheight=600truept]{geometry}
\usepackage{latexsym,amssymb,amsmath,graphicx,dsfont,slashed}
\usepackage[T1]{fontenc}

\def\theequation{\arabic{section}.\arabic{equation}}

\renewcommand{\theequation}{\thesection.\arabic{equation}}
\global\arraycolsep=1truept

\begin{document}

\hfill IFUP-TH/2010-47

\vskip 1.4truecm

\begin{center}
{\huge \textbf{Low-energy Phenomenology Of}}

\vskip .3truecm

{\huge \textbf{Scalarless Standard-Model Extensions }}

\vskip .4truecm

{\huge \textbf{With High-Energy Lorentz Violation }}

\vskip 1truecm

\textsl{Damiano Anselmi}$^{a,b}$\textsl{\ and Emilio Ciuffoli}$^{b,c}$

\vskip .2truecm

$^{a}$\textit{Institute of High Energy Physics, Chinese Academy of Sciences,}

\textit{19 (B) Yuquanlu, Shijingshanqu, Beijing 10049, China,}

\vskip .2truecm

$^{b}$\textit{Dipartimento di Fisica ``Enrico Fermi'', Universit\`{a} di
Pisa, }

\textit{Largo B. Pontecorvo 3, I-56127 Pisa, Italy,}

\vskip .2truecm

$^{c}$\textit{INFN, Sezione di Pisa, }

\textit{Largo B. Pontecorvo 3, I-56127 Pisa, Italy}

\vskip .2truecm

damiano.anselmi@df.unipi.it, emilio.ciuffoli@df.unipi.it

\vskip 1.5truecm

\textbf{Abstract}
\end{center}

\medskip

{\small We consider renormalizable Standard-Model extensions that violate
Lorentz symmetry at high energies, but preserve CPT, and do not contain
elementary scalar fields. A Nambu--Jona-Lasinio mechanism gives masses to
fermions and gauge bosons, and generates composite Higgs fields at low
energies. We study the effective potential at the leading order of the large-%
}$N_{c}${\small \ expansion, prove that there exists a broken phase and
study the phase space. In general, the minimum may break invariance under
boosts, rotations and CPT, but we give evidence that there exists a Lorentz
invariant phase. We study the spectrum of composite bosons and the
low-energy theory in the Lorentz phase. Our approach predicts relations
among the parameters of the low-energy theory. We find that such relations
are compatible with the experimental data, within theoretical errors. We
also study the mixing among generations, the emergence of the CKM\ matrix
and neutrino oscillations.}

\vskip 1truecm

\vfill\eject

\section{Introduction}

\setcounter{equation}{0}

Lorentz symmetry is a basic ingredient of the Standard Model of particle
physics and one of the best tested symmetries in nature \cite{kostelecky}.
From the theoretical viewpoint, we often take for granted that Lorentz
symmetry must be exact. However, our present knowledge cannot exclude that
it may just be approximate. Specifically, it could be violated at very high
energies or very large distances. Both possibilities have motivated several
authors to investigate the new physics that would emerge. Although no sign
of Lorentz violation has been found so far, these kinds of investigations
are useful because, after comparison with experiments, they allow us to put
bounds on the parameters of the violation. Among the most relevant reference
works, we mention refs. \cite{kostelecky2} and \cite{glashow}, together with
the data tables of \cite{kostelecky}. For a recent update on the state of
the art see ref. \cite{proceedings}.

In quantum field theory, if we assume that Lorentz symmetry is explicitly
violated at high energies we can turn non-renormalizable interactions into
renormalizable ones \cite{renolor}. In flat space, and in the realm of
perturbation theory, it is possible to construct gauge theories \cite
{LVgauge1suA,LVgauge1suAbar} and extensions of the Standard Model \cite
{lvsm,noH}, without violating physical principles. The Lorentz violating
models can contain several types of terms of higher dimensions. They are
multiplied by inverse powers of a scale $\Lambda _{L}$, which is interpreted
as the scale of Lorentz violation. Renormalizability holds because the
theory includes quadratic terms of higher dimensions that contain higher
space derivatives. The modified dispersion relations generate propagators
with improved ultraviolet behaviors. A ``weighted'' power counting,
according to which space and time have different weights, controls the
ultraviolet divergences of Feynman diagrams, and allows us to determine
which vertices are compatible with each set of quadratic terms. No terms
containing higher time derivatives (which would spoil unitarity) are
present, nor generated back by renormalization. Lorentz symmetry is
recovered at energies much smaller than $\Lambda _{L}$. It is not necessary
to assume that CPT\ is also violated to achieve these results.

The theories formulated using these tools are not meant to be just effective
field theories, but can be regarded as fundamental theories, in the sense
that, very much like the Standard Model, in principle they can describe
nature at arbitrarily high energies (when gravity is switched off). Some
Standard-Model extensions contain the vertex $(LH)^{2}/\Lambda _{L}$ at the
fundamental level, therefore give neutrinos Majorana masses after symmetry
breaking. No right-handed neutrinos, nor other extra fields, are necessary
to achieve this goal. Some extensions contain four fermion vertices $(\bar{%
\psi}\psi )^{2}/\Lambda _{L}^{2}$ at the fundamental level. Such vertices
can explain proton decay and trigger a Nambu--Jona-Lasinio mechanism, which
generates fermion masses, gauge-boson masses, and composite bosons. Finally,
some of our models can be phenomenologically viable even if they do not
contain elementary scalars.

Because of renormalizability, these extensions contain a finite set of
independent parameters, so they are to some extent predictive. It is
important to check that they can reproduce the known physics at low
energies. For example, it is not obvious that the models containing no
elementary scalars are able to fully reproduce the Standard Model at low
energies.

In this paper we study the low-energy phenomenology of the scalarless models
and compare predictions with experimental data. Although our approach has
large theoretical errors, the predictions are still meaningful, because they
can be falsified by data. We give enough evidence that our models do
reproduce the known low-energy physics. We study the effective potential in
detail, prove that there exists a broken phase, study the phase space and
give evidence that there exists a Lorentz phase. We investigate the mixing
among generations and show how the Cabibbo-Kobayashi-Maskawa (CKM)\ matrix
emerges. We study the spectrum of composite bosons that propagate at low
energies, the low-energy effective action, and the compatibility with
experimental constraints. Finally, we discuss neutrino oscillations.

The search for consistent Standard-Model alternatives that do not contain
elementary scalar fields has a long history, from technicolor \cite
{technicolor} to the more recent extra-dimensional Higgsless models \cite
{higghsless}. Worth mentioning are also the asymptotic-safety approach of
ref. \cite{wetterich} and the standard perturbative approach of ref. \cite
{slavnov}. In this respect, the violation of Lorentz symmetry offers, among
the other things, a new guideline and source of insight, and in our opinion
deserves the utmost attention.

The paper is organized as follows. In section 2 we review the minimal
scalarless Standard-Model extension that we are going to study. In section 3
we describe the dynamical symmetry-breaking mechanism and calculate the
effective potential to the leading order. In section 4 we prove that there
exists a domain in parameter space where the dynamical symmetry breaking
takes place, namely the effective potential has a non-trivial absolute
minimum. We also investigate the phase space. In section 5 we reconsider the
case of one generation treated in ref. \cite{noH} and prove some new
results. In section 6 we study the case of three generations and show how
the CKM\ matrix emerges. We also show that, in general, the Lorentz
violation predicts a more severe mixing among generations besides the CKM
matrix. In section 7 we show that there exist CPT violating local minima. In
section 8 we derive and study the low-energy effective action in the Lorentz
phase, in the case of one generation, and compare predictions with data. In
section 9 we show that the minimal model cannot generate (Majorana) masses
for left-handed neutrinos. Nevertheless, it is possible that neutrino
oscillations are explained in a different way. Section 10 contains
conclusions and outlook.

\section{The model}

\setcounter{equation}{0}

We assume that CPT and invariance under rotations are preserved. The
(minimal) scalarless model we are going to study reads 
\begin{equation}
\mathcal{L}_{\mathrm{noH}}=\mathcal{L}_{Q}+\mathcal{L}_{\text{kin}%
f}-\sum_{I=1}^{5}\frac{1}{\Lambda _{L}^{2}}g\bar{D}\bar{F}\,(\bar{\chi}_{I}%
\bar{\gamma}\chi _{I})+\frac{Y_{f}}{\Lambda _{L}^{2}}\bar{\chi}\chi \bar{\chi%
}\chi -\frac{g}{\Lambda _{L}^{2}}\bar{F}^{3},  \label{noH}
\end{equation}
where 
\begin{eqnarray}
\mathcal{L}_{\text{kin}f} &=&\sum_{a,b=1}^{3}\sum_{I=1}^{5}\bar{\chi}_{I}^{a}%
\hspace{0.02in}i\left( \delta ^{ab}\gamma ^{0}D_{0}-\frac{b_{0}^{Iab}}{%
\Lambda _{L}^{2}}{\bar{D}\!\!\!\!\slash}\,^{3}+b_{1}^{Iab}\bar{D}\!\!\!\!%
\slash \right) \chi _{I}^{b},  \nonumber \\
\mathcal{L}_{Q} &=&\frac{1}{4}\sum_{G}\left(
2F_{0i}^{G}F_{0i}^{G}-F_{ij}^{G}\tau ^{G}(\bar{\Upsilon})F_{ij}^{G}\right)
\label{kinkin}
\end{eqnarray}
are the quadratic terms of fermions and gauge fields, respectively, and $%
\Lambda _{L}$ is the scale of Lorentz violation. Bars are used to denote
space components and $\bar{F}$ denotes the ``magnetic'' components $F_{ij}$
of the field strengths. Moreover, $\chi _{1}^{a}=L^{a}=(\nu _{L}^{a},\ell
_{L}^{a})$, $\chi _{2}^{a}=Q_{L}^{a}=(u_{L}^{a},d_{L}^{a})$, $\chi
_{3}^{a}=\ell _{R}^{a}$, $\chi _{4}^{a}=u_{R}^{a}$ and $\chi
_{5}^{a}=d_{R}^{a}$, $\nu ^{a}=(\nu _{e},\nu _{\mu },\nu _{\tau })$, $\ell
^{a}=(e,\mu ,\tau )$, $u^{a}=(u,c,t)$ and $d^{a}=(d,s,b)$. The sum $\sum_{G}$
is over the gauge groups $SU(3)_{c}$, $SU(2)_{L}$ and $U(1)_{Y}$. The last
three terms of (\ref{noH}) are symbolic. Finally, $\bar{\Upsilon}\equiv -%
\bar{D}^{2}/\Lambda _{L}^{2}$ and $\tau ^{G}$ are polynomials of degree 2.
Gauge anomalies cancel out exactly as in the Standard Model \cite{lvsm}.

The model is ``minimal'' in the sense that it contains the minimal set of
elementary fields. It contains fewer fields than the minimal Standard Model,
because we have suppressed the elementary scalars. No right-handed
neutrinos, nor other extra fields, are included.

The model is renormalizable by weighted power counting in two ``weighted
dimensions''. This means that at high energies renormalizability is governed
by a power counting that resembles the one of a two dimensional field
theory, where energy has weight one, and the three space coordinates
altogether have weight one, therefore each of them separately has weight 1/3.

The weights of fields and couplings are determined so that each Lagrangian
term has weight 2. Gauge couplings $g$ have weight 1/3, so they are
super-renormalizable. For this reason, at very high energies gauge fields
become free and decouple, so the theory (\ref{noH}) becomes a four fermion
model in two weighted dimensions, described by the Lagrangian 
\begin{equation}
\mathcal{L}_{\mathrm{4f}}=\sum_{a,b=1}^{3}\sum_{I=1}^{5}\bar{\chi}_{I}^{a}%
\hspace{0.02in}i\left( \delta ^{ab}\gamma ^{0}\partial _{0}+b_{1}^{Iab}\bar{%
\partial}\!\!\!\!\hspace{1.1truept}\slash -\frac{b_{0}^{Iab}}{\Lambda _{L}^{2}}{\bar{\partial}%
\!\!\!\!\hspace{1.1truept}\slash}\,^{3}\right) \chi _{I}^{b}+\frac{Y_{f}}{\Lambda _{L}^{2}}\bar{%
\chi}\chi \bar{\chi}\chi .  \label{vnoH}
\end{equation}
We have kept also the terms multiplied by $b_{1}^{Iab}$, since they are
necessary to recover Lorentz invariance at low energies.

Our purpose is to investigate whether (\ref{noH}) can describe the known
low-energy physics by means of a dynamical symmetry breaking mechanism
triggered by four-fermion vertices.

The low-energy limit is the limit $\Lambda _{L}\rightarrow \infty $. From
the point of view of renormalization, power-like and logarithmic divergences
in $\Lambda _{L}$ appear in this limit, and add to the divergences already
present in the high-energy theory. The $\Lambda _{L}$-divergences make the
difference between the renormalization of the high-energy theory and the one
of the low-energy theory, which are controlled by weighted power counting
and ordinary power counting, respectively. When no symmetry-breaking
mechanism takes place, Lorentz symmetry can always be restored at low
energies fine-tuning the parameters of the low-energy Lagrangian. In ref. 
\cite{taiuti} these aspects of the low-energy limit have been studied in the
QED\ subsector of (\ref{noH}). However, taking the low-energy limit in the
full model (\ref{noH}) is more involved.

Because of the dynamical symmetry breaking mechanism, the symmetries of the
low energy theory depend on the vacuum. In turn, the vacuum depends on the
coefficients of the four-fermion vertices and the other free parameters of
the theory. The absolute minimum of the effective potential may break
boosts, and even rotations and CPT. If that happens, it is impossible to
recover Lorentz invariance at low energies and have compatibility with
experimental data. Thus, it is important to show that there exists a phase
(namely, a range in parameter space) where the minimum is Lorentz invariant,
so that Lorentz symmetry can be restored at low energies. One of the
purposes of this paper is to provide evidence that such a phase exists. This
is the phase where the Standard Model lives, and we call it the \textit{%
Lorentz phase}.

We proceed according to the following high-energy$\rightarrow $low-energy
pattern. It is useful to first switch gauge interactions off and switch them
back on later. Normally, this is just a trick to simplify the presentation,
but in our model it has a more physical justification, because, as explained
above, gauge fields decouple at very high energies, where the complete model
(\ref{noH}) reduces to the four-fermion model (\ref{vnoH}) plus free fields.
We show that the model (\ref{vnoH}) exhibits a dynamical symmetry-breaking
mechanism in the large $N_{c}$ expansion. Under suitable assumptions, we
argue that the effective potential has a Lorentz invariant minimum. The
minimum produces fermion condensates $\langle \bar{q}q\rangle $ and gives
masses to the fermions. Massive bound states (composite Higgs bosons)
emerge, together with Goldstone bosons. When gauge interactions are finally
switched back on, the Goldstone bosons associated with the breaking of $%
SU(2)_{L}\times U(1)_{Y}$ to $U(1)_{Q}$ are ``eaten'' by the $W^{\pm }$ and $%
Z$ bosons, which then become massive.

When we study the compatibility of our predictions with experimental data we
set the scale of Lorentz violation $\Lambda _{L}$ to $10^{14}$GeV. This
value was suggested in ref. \cite{lvsm} assuming that neutrino masses are
due to the vertex 
\begin{equation}
\frac{1}{\Lambda _{L}}(LH)^{2}.  \label{lh4}
\end{equation}
However, in the minimal model (\ref{noH}) this vertex is absent, both at the
fundamental and effective levels, and neutrino oscillations must be
explained in a different way (see section 9). Still, a number of
considerations suggest that $\sim 10^{14\text{-}15}$GeV are meaningful
values for the scale of Lorentz violation. They can be thought of as the
smallest values allowed by data. For example, they also agree with existing
bounds on proton decay, derived from four-fermion vertices $(\bar{\psi}\psi
)^{2}/\Lambda _{L}^{2}$: if we assume that the dimensionless coefficients
multiplying such vertices are of order one, we obtain $\Lambda _{L}\gtrsim
10^{15}$GeV \cite{astrumia}. For other, recent considerations on the
magnitude of $\Lambda _{L}$ and compatibility with ultrahigh-energy cosmic
rays, see \cite{taiuti2}.

\section{Dynamical symmetry-breaking mechanism}

\setcounter{equation}{0}

In this section we describe the dynamical symmetry breaking mechanism in the
model (\ref{vnoH}) and calculate the effective potential to the leading
order of the $1/N_{c}$ expansion.

The most general four-fermion vertices can be expressed using auxiliary
fields, that we call $M$, $N$, a quadratic potential $V_{2}$ and Yukawa
terms: 
\[
V_{2}(M,N)+\sum_{\alpha \beta ABIJ}\left[ M_{\alpha \beta ,IJ}^{AB}\bar{\psi}%
_{I}^{\alpha A}\psi _{J}^{\beta B}+\left( N_{\alpha \beta ,IJ}^{AB}\psi
_{I}^{\alpha A}\psi _{J}^{\beta B}+\text{H.c.}\right) \right] . 
\]
Here $\alpha ,\beta $ are spinor indices, $I,J$ are indices that denote the
type of fermions, $A,B$ are $SU(N_{c})\times SU(2)_{L}$-indices. $V_{2}(M,N)$
is the most general quadratic potential that is invariant under $%
SU(N_{c})\times SU(2)_{L}\times U(1)_{Y}$ and CPT. The Yukawa terms are made
symmetric assigning suitable transformation properties to the fields $M$ and 
$N$.

The four-fermion vertices are obtained integrating out the auxiliary fields $%
M$ and $N$. Several combinations of auxiliary fields may produce the same
four-fermion vertices. We do not need to select a minimal set of auxiliary
fields here. Actually, we include the maximal set of auxiliary fields,
because we want to study all possible intermediate channels. Some components
of the fields $M$ and $N$ become propagating at low energies (composite
bosons), others remain non-propagating also after the symmetry breaking.

\paragraph{Large $N_{c}$ expansion}

The Nambu--Jona-Lasinio dynamical symmetry-breaking mechanism is not
perturbative in the usual sense, so we need to have a form of control on it.
We use a large $N_{c}$ expansion. A rough estimate of the error due to the
large $N_{c}$ expansion can be obtained summing all powers of $1/N_{c}$ with
opposite signs, assuming that higher order contributions are of the same
magnitude (apart from the powers of $1/N_{c}$ in front of them). Thus,
calling ``1'' a generic quantity, its corrections are 
\begin{equation}
\pm \sum_{k=1}^{\infty }\frac{1}{N_{c}^{k}}=\pm \frac{1}{N_{c}-1}.
\label{error}
\end{equation}
For the purposes of this paper we just need to consider the leading order of
the $1/N_{c}$ expansion. For $N_{c}=3$, formula (\ref{error}) tells us that
we have a $\pm 50\%$ of error. Even if this error is large, some of our
predictions are enough precise to be possibly ruled out.

We cannot exclude that other symmetry-breaking mechanisms may take place in
the exact model, but we do not consider such possibilities here, because we
do not have a form of control on them such as the one provided by the large $%
N_{c}$ expansion.

The leading order of the $1/N_{c}$ expansion receives contributions only
from color-singlet fermion bilinears, on which we focus for the moment. We
consider the Yukawa terms 
\begin{eqnarray}
\mathcal{L}_{\text{Y}} &=&-\sum_{abmn}\left[ S_{mn}^{ab}(\bar{Q}%
_{R}^{am}Q_{L}^{bn})+\bar{S}_{nm}^{ba}(\bar{Q}_{L}^{am}Q_{R}^{bn})+H_{\mu
,mn}^{ab}(\bar{Q}_{L}^{am}\gamma ^{\mu }Q_{L}^{bn})\right.  \nonumber \\
&&\left. +K_{\mu ,mn}^{ab}(\bar{Q}_{R}^{am}\gamma ^{\mu }Q_{R}^{bn})+L_{\mu
\nu ,mn}^{ab}(\bar{Q}_{R}^{am}\sigma ^{\mu \nu }Q_{L}^{bn})+\bar{L}_{\mu \nu
,nm}^{ba}(\bar{Q}_{L}^{am}\sigma ^{\mu \nu }Q_{R}^{bn})\right] ,  \label{LY}
\end{eqnarray}
where $Q_{R}^{a}=(u_{R}^{a},d_{R}^{a})$ and $m,n$ are both $SU(2)_{L}$- and $%
SU(2)_{R}$-indices, depending on the case. The Yukawa terms are $%
U(2)_{L}\times U(2)_{R}$-invariant, and so is the leading-order correction
to the effective potential. The (contracted) $SU(N_{c})$-indices are not
shown. The fields $S$ and $L_{\mu \nu }$ are CPT even, while the fields $%
H_{\mu }$ and $K_{\mu }$ are CPT odd. The matrices $H_{\mu }$ and $K_{\mu }$
are Hermitian.

\paragraph{Lagrangian of the high-energy model and effective potential}

As usual, we first switch the gauge fields off, because they decouple at
high energies. We will turn them back on later. The fermionic kinetic terms
are 
\[
\mathcal{L}_{\text{kf}}=\sum_{abm}\bar{Q}_{L}^{am}\hspace{0.02in}i\left(
\delta ^{ab}\gamma ^{0}\partial _{0}+b_{1L}^{ab}{\bar{\partial}\!\!\!\!\hspace{1.1truept}\slash}-%
\frac{b_{0L}^{ab}}{\Lambda _{L}^{2}}{\bar{\partial}\!\!\!\!\hspace{1.1truept}\slash}%
\,^{3}\right) Q_{L}^{bm}+\bar{Q}_{R}^{am}\hspace{0.02in}i\left( \delta
^{ab}\gamma ^{0}\partial _{0}+b_{1R}^{abm}{\bar{\partial}\!\!\!\!\hspace{1.1truept}\slash}-\frac{%
b_{0R}^{abm}}{\Lambda _{L}^{2}}{\bar{\partial}\!\!\!\!\hspace{1.1truept}\slash}\,^{3}\right)
Q_{R}^{bm}, 
\]
where $b_{0,1L}^{ab}$ and $b_{0,1R}^{abm}$ are Hermitian matrices for every $%
m$. The total Lagrangian reads 
\[
\mathcal{L}=\mathcal{L}_{\text{kf}}+\mathcal{L}_{\text{Y}}+\mathcal{L}_{%
\text{Y}}^{\prime }+W_{2}^{\prime }(S,H,K,L,N^{\prime }), 
\]
where $\mathcal{L}_{\text{Y}}^{\prime }$ and $N^{\prime }$ denote all other
Yukawa terms and auxiliary fields, respectively. The potential $%
W_{2}^{\prime }$ is the most general quadratic form compatible with the
symmetries of the theory. We can eliminate the off-diagonal terms $%
SN^{\prime }$, $HN^{\prime }$, $KN^{\prime }$ and $LN^{\prime }$ translating
the fields $N^{\prime }$. Calling $N$ the translated fields, we get a
quadratic potential of the form 
\[
W_{2}^{\prime }=W_{2}(S,H,K,L)+W_{2}^{\prime \prime }(N). 
\]
The leading-order correction to the potential depends only on $S$, $H$, $K$
and $L$. The fields $N$ have vanishing expectation values, so the fields $%
N^{\prime }$ can have non-trivial expectation values because of the
translation from $N^{\prime }$ to $N$. For the moment we can ignore the $N$%
-sector and focus on $W_{2}$.

By CPT and rotational invariance, the potential $W_{2}$ has the symbolic
structure 
\[
W_{2}(S,H,K,L)\sim SS^{\dagger
}+H_{0}^{2}+H_{i}^{2}+H_{0}K_{0}+H_{i}K_{i}+K_{0}^{2}+K_{i}^{2}+L_{0j}L_{0j}^{\dagger }+L_{ij}L_{ij}^{\dagger }. 
\]
The indices not shown explicitly in this formula are contracted with
constant tensors.

To study the effective potential we need to consider the Lagrangian 
\begin{equation}
\mathcal{L}_{q}=\sum_{a,b=1}^{3}\bar{\Psi}^{a}\left( i\Gamma ^{0}\mathds{1}%
\partial _{t}+i\bar{\Gamma}\cdot {\bar{\partial}}\left( B_{1}-\frac{{\bar{%
\partial}}^{2}}{\Lambda _{L}^{2}}B_{0}\right) -M\right) ^{ab}\Psi
^{b}-W_{2}(M),  \label{model2}
\end{equation}
where 
\[
B_{0,1}=\left( 
\begin{tabular}{cc}
$b_{0,1L}^{ab}\delta ^{mn}$ & $0$ \\ 
$0$ & $b_{0,1R}^{abp}\delta ^{pq}$%
\end{tabular}
\right) ,\qquad M=\left( 
\begin{tabular}{cc}
$S+L_{\mu \nu }\hat{\sigma}^{\mu \nu }$ & $K_{\mu }\sigma ^{\mu }$ \\ 
$H_{\mu }\bar{\sigma}^{\mu }$ & $S^{\dagger }+L_{\mu \nu }^{\dagger }\check{%
\sigma}^{\mu \nu }$%
\end{tabular}
\right) , 
\]
and 
\[
(\Gamma ^{\mu })^{ab}=\delta ^{ab}\left( 
\begin{tabular}{cc}
$0$ & $\Sigma ^{\mu }$ \\ 
$\bar{\Sigma}^{\mu }$ & $0$%
\end{tabular}
\right) ,\quad \Psi ^{a}=\left( 
\begin{tabular}{c}
$Q_{L}^{am}$ \\ 
$Q_{R}^{ap}$%
\end{tabular}
\right) , 
\]
$(\Sigma ^{\mu })^{mp}=\delta ^{mp}\sigma ^{\mu }$, $(\bar{\Sigma}^{\mu
})^{pm}=\delta ^{mp}\bar{\sigma}^{\mu }$, $\sigma ^{\mu }=(1, %
\mbox{\boldmath$\sigma$} )$, $\bar{\sigma}^{\mu }=(1,-\mbox{\boldmath$%
\sigma$} )$, $\check{\sigma}^{\mu \nu }=-i(\bar{\sigma}^{\mu }\sigma ^{\nu }-%
\bar{\sigma}^{\nu }\sigma ^{\mu })/2$, $\hat{\sigma}^{\mu \nu }=-i(\sigma
^{\mu }\bar{\sigma}^{\nu }-\sigma ^{\nu }\bar{\sigma}^{\mu })/2$.

The leading-order effective potential reads 
\[
W(M)=W_{2}(M)+\mathcal{V}(M), 
\]
where $\mathcal{V}(M)$ is calculated integrating over the fermions. It is
the renormalized version of 
\[
\mathcal{V}_{\text{div}}(M)\equiv -N_{c}\int^{\Lambda }\frac{\text{\textrm{d}%
}^{4}p}{(2\pi )^{4}}\ln \det \left( P-\Gamma ^{0}M\right) ,\qquad P\equiv i%
\mathds{1}p_{4}+\Gamma ^{0}\bar{\Gamma}\cdot \bar{p}\left( B_{1}+\frac{\bar{p%
}^{2}}{\Lambda _{L}^{2}}B_{0}\right) . 
\]
The integral has already been rotated to the Euclidean space. We regulate
the ultraviolet divergences with a cut-off $\Lambda $ and subtract them
expanding in $M$ around $M=0$. The lowest order in $M$ is a constant, while
the first order in $M$ is proportional to the integral of tr$\left[
P^{-1}\Gamma ^{0}M\right] $, which is odd in momentum, so it vanishes. The
second order is logarithmically divergent, while all other orders are
convergent. We have 
\begin{equation}
\mathcal{V}(M)=-N_{c}\int \frac{\text{\textrm{d}}^{4}p}{(2\pi )^{4}}\left(
\ln \det (\mathds{1}-P^{-1}\Gamma ^{0}M)+\frac{1}{2}\text{tr}\left[
P^{-1}\Gamma ^{0}MP^{-1}\Gamma ^{0}M\right] \right) .  \label{vm}
\end{equation}
Observe that $\mathcal{V}(M)$ is regular in the infrared.

\section{Existence of a non-trivial absolute minimum}

\setcounter{equation}{0}

In this section we prove that there exists a phase where the dynamical
symmetry-breaking mechanism takes place. Precisely, the potential has a
nontrivial absolute minimum if some parameters contained in $W_{2}(M)$
satisfy certain bounds and $B_{1}$ is in the neighborhood of the identity.
The assumption $B_{1}\sim 1$ is not only useful to simplify the
calculations, but also justified by all known experimental data \cite
{kostelecky}.

It is sufficient to work at $B_{1}=1$, because the result, once proved for $%
B_{1}=1$, extends to the neighborhood of the identity by continuity. On the
other hand, for the moment we keep the matrix $B_{0}$ free, because its
entries can differ from one another by several orders of magnitude.

We first prove that the potential grows for large $M$, in all directions.
This result allows us to conclude that there exists an absolute minimum.
Indeed, since the function $W(M)$ is continuous the extreme value theorem
ensures that it has absolute maxima and absolute minima in an arbitrary
sphere $|M|\leqslant R$. If we take $R$ large enough $W(M)$ grows outside
the sphere. Then the absolute minima inside the sphere are absolute minima
of the function.

Later we show that, in a suitable domain $\mathcal{D}$ of parameter space,
the point $M=0$, which is stationary, is not a minimum. This proves that the
absolute minimum of $W(M)$ is nontrivial in $\mathcal{D}$. Along with the
proof, we derive the bounds that define $\mathcal{D}$.

To study $W(M)$ for large $M$, we rescale $M$ by a factor $\lambda $ and
then let $\lambda $ tend to infinity. It is useful to rescale also $p_{4}$
by a factor $\lambda $ and $\bar{p}$ by a factor $\lambda ^{1/3}$. We get 
\begin{equation}
W(\lambda M)=-\frac{N_{c}\lambda ^{2}}{2}\int_{-\infty }^{+\infty }\frac{%
\text{\textrm{d}}p_{4}}{2\pi }\int_{\text{IR}}\frac{\text{\textrm{d}}^{3}%
\bar{p}}{(2\pi )^{3}}\text{tr}\left[ \hat{P}^{-1}\Gamma ^{0}M\hat{P}%
^{-1}\Gamma ^{0}M\right] +\mathcal{O}(\lambda ^{2}),  \label{winte}
\end{equation}
where $\hat{P}$ is the same as $P$, but with $B_{1}\rightarrow B_{1}/\lambda
^{2/3}$. The subscript IR means that the $\bar{p}$-integral is restricted to
the IR region. It gives contributions proportional to $\lambda ^{2}\ln
\lambda ^{2}$.

Formula (\ref{winte}) is proved as follows. If we factor out a $\lambda ^{2}$
and take $\lambda $ to infinity inside the integrand of (\ref{vm}), we
notice that the integral remains convergent in the ultraviolet region, but
becomes divergent in the infrared region. Thus, when $\lambda \rightarrow
\infty $ the infrared region provides dominant contributions that grow
faster than $\lambda ^{2}$. The first term of (\ref{vm}) does not give
dominant contributions: indeed, in the IR region it is safe to take $\lambda
\rightarrow \infty $ inside the logarithm. Instead, it is not safe to do the
same in the second term of (\ref{vm}). This explains formula (\ref{winte}).

Now we calculate the dominant contributions. It is convenient to work in the
basis where the matrix $B_{0}$ is diagonal: 
\begin{equation}
B_{0}^{ab}=\delta ^{ab}\text{diag}\left(
b_{L}^{a},b_{L}^{a},b_{uR}^{a},b_{dR}^{a}\right) .  \label{basisdiag}
\end{equation}
Here the indices $u$,$d$ refer to the ``up'' and ``down'' quarks of the
family labeled by the index $a$ (so they mean $c$,$s$ and $t$,$b$ for $a=2$
and $3$, respectively). In this basis the propagator is diagonal in $a$,$b$.
The trace is invariant under rotations, so it can be calculated orienting $%
\bar{p}$ along the $z$-direction and rewriting the result as a scalar. With
this choice, the propagator is diagonal in all indices, and the trace can be
easily calculated. We obtain a linear combination of integrals of the form 
\[
\int_{-\infty }^{+\infty }\frac{\text{\textrm{d}}p_{4}}{2\pi }\int_{\text{IR}%
}\frac{\text{\textrm{d}}^{3}\bar{p}}{(2\pi )^{3}}\frac{1}{ip_{4}+X}\frac{1}{%
ip_{4}+Y}. 
\]
The integral in $p_{4}$ can be calculated using the residue theorem. The $%
\bar{p}$-integrand, which is quadratic in $M$, is at most quadratic in the
components of $\bar{p}$, and can be symmetrized using 
\[
\bar{p}^{i}\bar{p}^{j}\rightarrow \frac{\delta _{ij}}{3}\bar{p}^{2}. 
\]
We obtain a linear combination of $\bar{p}$-integrals of the form 
\[
\int_{\text{IR}}\frac{\text{\textrm{d}}^{3}\bar{p}}{(2\pi )^{3}}\frac{1}{|%
\bar{p}|\left( \frac{1}{\lambda ^{2/3}}+\frac{\bar{p}^{2}}{\Lambda _{L}^{2}}%
b_{xy}\right) }\sim \frac{\Lambda _{L}^{2}}{3(2\pi )^{2}b_{xy}}\ln \lambda
^{2}, 
\]
where $b_{xy}$ is the sum of two entries of the matrix $B_{0}$. The result
is a linear combination of contributions of the form 
\begin{equation}
N_{c}\Lambda _{L}^{2}c_{xy}\frac{|(\Gamma ^{0}M)_{xy}|^{2}}{b_{xy}}\lambda
^{2}\ln \lambda ^{2},  \label{defpo}
\end{equation}
where $c_{xy}$ is a non-negative numerical factor. Converting the result to
a generic basis, where $B_{0}$ is not necessarily diagonal, we find 
\[
W(\lambda M)=W_{\text{dom}}(\lambda M)+\mathcal{O}(\lambda ^{2}), 
\]
with 
\begin{eqnarray}
W_{\text{dom}}(\lambda M) &=&\frac{N_{c}\Lambda _{L}^{2}}{6(2\pi )^{2}}%
\lambda ^{2}\ln \lambda ^{2}\int_{0}^{\infty }\mathrm{d}\xi \hspace{0.01in}%
\text{tr}\left[ \mathcal{S}\mathrm{e}^{-\xi B_{0}}\mathcal{S}\mathrm{e}%
^{-\xi B_{0}}+\frac{2}{3}\mathcal{H}_{i}\mathrm{e}^{-\xi B_{0}}\mathcal{H}%
_{i}\mathrm{e}^{-\xi B_{0}}\right.  \nonumber \\
&&\qquad \qquad \qquad \qquad \left. +\frac{2}{3}\mathcal{K}_{i}\mathrm{e}%
^{-\xi B_{0}}\mathcal{K}_{i}\mathrm{e}^{-\xi B_{0}}+\frac{1}{3}\mathcal{G}%
_{i}\mathrm{e}^{-\xi B_{0}}\mathcal{G}_{i}\mathrm{e}^{-\xi B_{0}}\right] ,
\label{reso}
\end{eqnarray}
where $\mathcal{S}$, $\mathcal{H}_{i}$, $\mathcal{K}_{i}$, $\mathcal{G}_{i}$
are matrices obtained from $\Gamma ^{0}M$ dropping all entries that are not $%
S$, $H_{i}$, $K_{i}$ and $G_{i}\equiv 2iL_{0i}-\varepsilon _{ijk}L_{jk}$,
respectively.

The dominant contribution (\ref{reso}) of $W(\lambda M)$ is
positive-definite in the $M$-entries that it contains. Indeed, recalling
that $B_{0}$ and $\Gamma ^{0}M$ are Hermitian, the integrand is the sum of
terms of the form 
\[
\text{tr}\left[ \left( \mathrm{e}^{-\xi B_{0}/2}\mathcal{M}\mathrm{e}^{-\xi
B_{0}/2}\right) \left( \mathrm{e}^{-\xi B_{0}/2}\mathcal{M}\mathrm{e}^{-\xi
B_{0}/2}\right) ^{\dagger }\right] , 
\]
which are positive definite. Thus, the effective potential grows in all
directions on which $W_{\text{dom}}(\lambda M)$ depends.

However, $W_{\text{dom}}(\lambda M)$ does not depend on all $M$-entries.
Precisely, it does not contain $H_{0}$, $K_{0}$ and $L_{0i}$ (in the basis $%
L_{0i}$-$G_{i}$). Thus, the dominant contributions of $\mathcal{V}(\lambda
M) $ that depend on such entries are at most of order $\lambda ^{2}$, as are
the contributions coming from the tree-level potential $W_{2}(\lambda M)$.
Now, $\mathcal{V}(\lambda M)$ is uniquely determined, while $W_{2}(\lambda
M) $ contains free parameters. If we assume that the $W_{2}(\lambda M)$%
-coefficients that multiply the terms containing $H_{0}$, $K_{0}$ and $%
L_{0i} $ satisfy suitable inequalities, which define a certain domain $%
\mathcal{D}^{\prime }$ in parameter space, the total leading-order potential 
$W(\lambda M)$ grows in all directions. Then, by continuity, it must have a
minimum somewhere. This is not the end of our argument, since the minimum
could still be trivial.

Let us investigate the point $M=0$. It is certainly a stationary point,
since the first derivatives of both $W_{2}(M)$ and $\mathcal{V}(M)$ vanish
at $M=0$. Moreover, the second derivatives of $\mathcal{V}(M)$ vanish at $%
M=0 $ by construction, so the second derivatives of $W(M)$ at $M=0$ coincide
with those of $W_{2}(M)$. Thus, choosing some free parameters of $W_{2}(M)$
to be negative, or smaller that certain bounds, we can define a domain $%
\mathcal{D}^{\prime \prime }$ in parameter space where the origin $M=0$ is
not a local minimum.

The domains $\mathcal{D}^{\prime }$ and $\mathcal{D}^{\prime \prime }$ have
a non-empty intersection $\mathcal{D}$. Indeed, it is sufficient to choose a 
$\mathcal{D}^{\prime \prime}$-region defined by bounds on the $W_{2}(M)$%
-parameters that are unrelated to $H_{0}$, $K_{0}$ and $L_{0i}$.

In the domain $\mathcal{D}$ the potential $W(M)$ grows in every direction
for large $M$, therefore it has a minimum. Moreover, the minimum cannot be
the origin, but it is located somewhere at $M\neq 0$. This means that the
symmetry-breaking mechanism necessarily takes place in $\mathcal{D}$, as we
wanted to prove.

\paragraph{Phase diagram}

Varying the parameters contained in $W_{2}$, the absolute minimum moves
around and we can study the phase diagram of the theory.

So far, we have rigorously proved that the theory has an unbroken phase and
a broken phase. We still do not know much about the minimum of the broken
phase. To make contact with experiments it is necessary to prove that there
exists a broken phase that $i$) preserves rotations and CPT and $ii$) allows
us to recover Lorentz symmetry at low energies. In this Lorentz phase only
the fields $S$ may have non-trivial expectation values, while $H_{\mu }$, $%
K_{\mu }$ and $L_{\mu \nu }$ must vanish at the minimum.

A number of technical difficulties prevent us from rigorously prove that the
Lorentz phase exists in the most general case. However, we give a number of
results providing evidence that it does exist in several particular cases of
interest.

Le us assume for the moment that tuning the $W_{2}(M)$-parameters we can
obtain every configuration of expectation values we want. Then the theory
has a rich phase diagram. Besides the unbroken phase and\ the Lorentz phase,
we have broken phases where Lorentz symmetry is violated also at low
energies, namely some vector fields or tensor fields acquire non-trivial
expectation values. Among these phases, we have: $i$) a phase where
invariance under rotations is preserved, but CPT is broken, if $%
H_{i}=K_{i}=L_{\mu \nu }=0$ at the minimum and $H_{0}$, $K_{0}$ have
non-trivial expectation values; $ii$) a phase where rotational invariance is
broken, but CPT\ is preserved, if $H_{\mu }=K_{\mu }=0$, but $L_{\mu \nu
}\neq 0$; $iii$) a phase where rotational invariance and CPT are both
broken, if $H_{i}$, $K_{i}$ have non-trivial expectation values. Note that
there is no Lorentz violating phase where CPT\ and invariance under
rotations are both preserved.

At the leading order of the $1/N_{c}$ expansion it is consistent to project
onto the scalar sector putting $H_{\mu }=K_{\mu }=L_{\mu \nu }\equiv 0$,
because such fields are generated back by renormalization only at subleading
orders. Equivalently, adding quadratic terms proportional to $H^{2}$, $K^{2}$
and $L^{2}$ to the tree-level potential $W_{2}(M)$, multiplied by
arbitrarily large positive coefficients, it is possible to freeze the vector
and tensor directions at the leading order. Then, the expectation values of $%
H_{\mu }$, $K_{\mu }$ and $L_{\mu \nu }$ become arbitrarily small and may be
assumed to be zero for all practical purposes. This argument can partially
justify the existence of the Lorentz phase and the projection onto the
scalar subsector, which we advocate in the next sections. However, we stress
that it works only at the leading order of the $1/N_{c}$ expansion.

\subsection{Lorentz invariant local minimum}

We begin to study the Lorentz phase investigating when the point 
\begin{equation}
S=S_{0}\neq 0,\qquad H_{\mu }=K_{\mu }=L_{\mu \nu }=0  \label{locmin}
\end{equation}
is a local minimum. Again, we consider a neighborhood of $B_{1}=1$ (which
allows us to work at precisely $B_{1}=1$, by continuity) and restrict the
tree-level couplings of $W_{2}(M)$ to a suitable domain in parameter space.

Consider the first derivatives $\partial W/\partial M$ calculated at (\ref
{locmin}). Clearly, both $\partial W/\partial H$ and $\partial W/\partial K$
vanish, since they are CPT odd, and $\partial W/\partial L_{0i}$ and $%
\partial W/\partial L_{ij}$ vanish by invariance under rotations. Instead, $%
\partial W/\partial S=\partial W_{2}/\partial S+\partial \mathcal{V}%
/\partial S$ can be made to vanish adjusting the free parameters that
multiply the $S$-$\bar{S}$-quadratic terms contained in $W_{2}$. Observe
that all other $W_{2}$-parameters remain arbitrary, a fact that will be
useful in a moment.

Now we study the second derivatives $\partial ^{2}W/\partial M^{2}$ at the
point (\ref{locmin}). Assume that the matrix 
\begin{equation}
\left( 
\begin{tabular}{cc}
$\frac{\partial ^{2}W}{\partial S^{2}}$ & $\frac{\partial ^{2}W}{\partial
S\partial \bar{S}}$ \\ 
$\frac{\partial ^{2}W}{\partial \bar{S}\partial S}$ & $\frac{\partial ^{2}W}{%
\partial \bar{S}^{2}}$%
\end{tabular}
\right)  \label{block}
\end{equation}
is positive definite at the minimum. The derivatives $\partial
^{2}W/(\partial H\partial S)$ and $\partial ^{2}W/(\partial K\partial S)$
vanish, since they are CPT odd. The derivatives $\partial ^{2}W/(\partial
S\partial L_{0i})$ and $\partial ^{2}W/(\partial S\partial L_{ij})$ vanish
by rotational invariance. The matrix $\partial ^{2}W/\partial M^{2}$ is then
block diagonal. One block is (\ref{block}) and the second block does not
contain derivatives with respect to $S$. The second block can be made
positive definite assuming that the $W_{2}$-parameters that have remained
arbitrary satisfy suitable inequalities.

We still have to prove that (\ref{block}) is positive definite. This
calculation is rather involved in a generic situation. We study this problem
in a number of special cases.

\section{The case of one generation revisited}

\setcounter{equation}{0}

While experiments tell us that the matrix $B_{1}$ is close to the identity,
we have no such information about the matrix $B_{0}$. Actually, its entries
could differ from one another by several orders of magnitude, so in
principle the matrix $B_{0}$ should be kept generic. However, calculations
with a generic $B_{0}$ are rather involved, so we have to make some
simplifying assumptions. In this section we reconsider the case of one
generation (which we assume to be the third one, for future use) in the
scalar sector with $B_{0}=B_{1}=1$ \cite{noH}. We also prove some statements
that were not proved in \cite{noH}, for example that the minimum is absolute
and unique in the scalar sector.

In the scalar sector $H_{\mu }=K_{\mu }=L_{\mu \nu }=0$. We have 
\[
\Gamma ^{0}M=\left( 
\begin{tabular}{cc}
$0$ & $\tau ^{\dagger }$ \\ 
$\tau $ & $0$%
\end{tabular}
\right) , 
\]
where $\tau $ is a $2\times 2$ matrix, with indices of $SU(2)_{L}$ to the
right and indices of $SU(2)_{R}$ to the left. The fermions are organized as $%
\Psi =(Q_{L},Q_{R})$ and $Q=(t,b)$.

If we assume the axial symmetry $U(1)_{A}$, besides $SU(2)_{L}$ and $%
U(1)_{Y} $, the leading-order potential is 
\begin{equation}
W(M)=\Lambda _{L}^{2}\text{tr}[\tau \tau ^{\dagger }C]-N_{c}\int \frac{\text{%
\textrm{d}}^{4}p}{(2\pi )^{4}}\left( \ln \det (\mathds{1}-P^{-1}\Gamma
^{0}M)+\frac{1}{2}\text{tr}\left[ P^{-1}\Gamma ^{0}MP^{-1}\Gamma
^{0}M\right] \right) ,  \label{wm}
\end{equation}
where $C$ is a diagonal constant matrix, $C=\hspace{0.03in}$diag$%
(c_{t},c_{b})$.

We use the ``polar'' decomposition (\ref{su}) to write 
\[
\tau =\tilde{U}_{R}DU_{L},\qquad D=\left( 
\begin{tabular}{cc}
$d_{t}$ & $0$ \\ 
$0$ & $d_{b}$%
\end{tabular}
\right) , 
\]
and the diagonalization (\ref{diago}) for $N=\Gamma ^{0}M$. See Appendix A
for notation and details. At $B_{0}=B_{1}=1$ the one-loop correction to the
potential does not depend on the diagonalizing matrix $U$ of (\ref{diago}),
but only on the entries $d_{t}$, $d_{b}$ of $D$. It is useful to define the
four vector 
\begin{equation}
p^{\prime }=\left( p^{0},\bar{p}\left( 1+\frac{\bar{p}^{2}}{\Lambda _{L}^{2}}%
\right) \right) ,  \label{4v}
\end{equation}
because the integrand of (\ref{wm}) is ``Lorentz invariant'' in this
four-vector, therefore it can be calculated at $\bar{p}=0$. Writing 
\[
\tilde{U}_{R}=\sqrt{1-|u|^{2}}+iu\sigma _{+}+i\bar{u}\sigma _{-}, 
\]
where $\sigma _{\pm }=(\sigma _{1}\pm i\sigma _{2})/2$, $|u|\leqslant 1$, we
obtain the potential 
\[
W(M)=\Lambda _{L}^{2}(d_{t}^{2}c_{t}+d_{b}^{2}c_{b})-\Lambda
_{L}^{2}|u|^{2}(d_{t}^{2}-d_{b}^{2})(c_{t}-c_{b})+2V(d_{t}^{2})+2V(d_{b}^{2}), 
\]
where 
\[
V(r)\equiv -N_{c}\int \frac{\text{\textrm{d}}^{4}p}{(2\pi )^{4}}\left( \ln
\left( 1+\frac{r}{(p^{\prime })^{2}}\right) -\frac{r}{(p^{\prime })^{2}}%
\right) . 
\]
This function is non-negative, monotonically increasing and convex. Indeed,
for $r>0$, 
\[
V^{\prime }(r)=rN_{c}\int \frac{\text{\textrm{d}}^{4}p}{(2\pi )^{4}}\frac{1}{%
(p^{\prime })^{2}((p^{\prime })^{2}+r)}>0,\qquad V^{\prime \prime
}(r)=N_{c}\int \frac{\text{\textrm{d}}^{4}p}{(2\pi )^{4}}\frac{1}{%
((p^{\prime })^{2}+r)^{2}}>0. 
\]
Moreover, $V(0)=V^{\prime }(0)=0$ and $V^{\prime \prime }(0)=+\infty $.

Let us find the stationary points of $W(M)$ and study the Hessians there. We
denote the values of $d_{t,b}$ at the stationary point with $m_{t,b}$, and
identify them with the top and bottom masses, respectively.

We assume $c_{t}\neq c_{b}$, because, as we prove below, the case $%
c_{t}=c_{b}$ is not physically interesting. Then we find the following
stationary points:

\noindent 1) $u=0$, while $m_{t,b}$ do not vanish and solve the gap equation 
\begin{equation}
\Lambda _{L}^{2}c_{i}+2V^{\prime }(m_{i}^{2})=0;  \label{gap}
\end{equation}
2) $u=0$, while one of $m_{t,b}$ vanishes and the other one solves the gap
equation (\ref{gap});

\noindent 3)$\;|u|^{2}=1/2$ and $m_{t}=m_{b}\neq 0$ solve 
\begin{equation}
0=\Lambda _{L}^{2}(c_{t}+c_{b})+4V^{\prime }(m_{t}^{2}).  \label{gap2}
\end{equation}
4) $m_{t}=m_{b}=0$.

Now we analyze the Hessian at each stationary point.

\noindent 1) Because of (\ref{gap}) and $c_{t}\neq c_{b}$, and since $%
V^{\prime }$ is monotonic, $m_{t}$ and $m_{b}$ cannot coincide. The Hessian
is diagonal and strictly positive: 
\[
\left. \frac{\partial ^{2}W}{\partial d_{i}^{2}}\right| _{\text{min}%
}=8m_{i}^{2}V^{\prime \prime }(m_{i}^{2})>0,\qquad \left. \frac{\partial
^{2}W}{\partial |u|^{2}}\right| _{\text{min}}=2(m_{t}^{2}-m_{b}^{2})(V^{%
\prime }(m_{t}^{2})-V^{\prime }(m_{b}^{2}))>0. 
\]
\noindent This point is a local minimum. It exists if and only if the gap
equations (\ref{gap}) have solutions, which occurs if and only if both $%
c_{t} $ and $c_{b}$ are negative.

\noindent 2) If $m_{b}$ vanishes then the Hessian is diagonal and 
\[
\left. \frac{\partial ^{2}W}{\partial d_{t}^{2}}\right| _{\text{min}%
}=8m_{t}^{2}V^{\prime \prime }(m_{t}^{2}),\qquad \left. \frac{\partial ^{2}W%
}{\partial d_{b}^{2}}\right| _{\text{min}}=2\Lambda _{L}^{2}c_{b},\qquad
\left. \frac{\partial ^{2}W}{\partial |u|^{2}}\right| _{\text{min}}=\Lambda
_{L}^{2}m_{t}^{2}(c_{b}-c_{t}). 
\]
\noindent This point is a local minimum if and only if 
\[
c_{b}>0,\qquad c_{b}>c_{t}. 
\]

\noindent 3) The determinant of the Hessian is negative, 
\[
\det H=-32\Lambda _{L}^{4}m^{4}(c_{t}-c_{b})^{2}V^{\prime \prime }(m^{2}), 
\]
so this point cannot be the minimum.

\noindent 4) From the analysis of the previous section we already know that
the origin is a local minimum if and only if both $c_{i}$'s are positive.

The physically interesting case is clearly 1). Since both $c_{t}$ and $c_{b}$
are negative, we may assume 
\begin{equation}
c_{t}<c_{b}<0.  \label{condition}
\end{equation}
Then point 1) is the unique local minimum in the scalar sector. The theorem
proved in the previous section (existence of the absolute minimum) allows us
to conclude that point 1) is also the absolute minimum of $W(M)$ in the
scalar sector. Moreover, the argument of section 4.1 ensures that if the
other tree-level couplings of $W_{2}(M)$ belong to a suitable region in
parameter space point 1) is also a local minimum in the full $M$-space.

Note that these arguments still do not prove that there exists a phase where
point 1) is the absolute minimum in the full $M$-space.

Because of the symmetries of the potential, its minimum is not just a point,
but a geometric locus of points. By means of a $SU(2)_{L}\times
U(1)_{Y}\times U(1)_{A}$-transformation we can choose the physical minimum 
\begin{equation}
\tau _{0}=\left( 
\begin{tabular}{cc}
$m_{t}$ & $0$ \\ 
$0$ & $m_{b}$%
\end{tabular}
\right) ,  \label{desi}
\end{equation}
which preserves $U(1)_{Q}$.

The other cases are not physically interesting. For example, if either $%
c_{t} $ or $c_{b}$ vanish or are positive the absolute minimum is either
point 2)\ or the origin $M=0$. Then at least one mass vanishes. Instead, if $%
c_{t}=c_{b}$ the theory is invariant under the custodial symmetry $SU(2)_{R}$
and $m_{t,b}$ either vanish or solve the gap equation (\ref{gap}). Using $%
SU(2)_{R}\times SU(2)_{L}\times U(1)_{Y}\times U(1)_{A}$ we can always make
the minimum have the form (\ref{desi}), but either some masses vanish or
coincide.

We conclude that there is a (unique, up to exchange of $m_{t}$ and $m_{b}$)
phase such that $W(M)$ has the absolute minimum (\ref{desi}) in the scalar
sector, and point (\ref{desi}) is also a local minimum in the full $M$-space.

\section{Three generations}

\setcounter{equation}{0}

Now we study the case of three generations, focusing again on the scalar
sector and still assuming $B_{0}=B_{1}=1$. We look for evidence that the
Lorentz phase exists. Assuming again axial symmetry, the potential $%
W(M)=W_{2}(M)+\mathcal{V}(M)$ has 
\begin{equation}
W_{2}(M)=\Lambda _{L}^{2}\sum_{mnabcd}S_{mn}^{ab}\bar{S}%
_{mn}^{cd}C_{m}^{abcd},\qquad \mathcal{V}(M)=2\sum_{i}V(d_{i}^{2}),
\label{mosto}
\end{equation}
where $C_{m}^{abcd}$ are constants. The correction $\mathcal{V}(M)$ is
calculated using the polar decomposition (\ref{diago}) for $N=\Gamma ^{0}M$
and noting that the integrand is independent of $U$. Moreover, it is Lorentz
invariant in the four-vector (\ref{4v}), so it can be easily calculated at $%
\bar{p}=0$ and later rewritten in covariant form.

As before, $\mathcal{V}(M)$ is positive definite, monotonically increasing
and convex. Its minimum is $M=0$, so the minimum of $W(M)$ is determined by
the free parameters $C_{m}^{abcd}$ contained in $W_{2}(M)$.


\paragraph{Illustrative example}

To begin with, it is worth considering the simple case 
\begin{equation}
C_{m}^{abcd}=H^{bd}C_{m}^{ca},  \label{asso}
\end{equation}
where $H$ and $C_{m}$ are Hermitian matrices.

Define the matrices $\mathcal{H}_{nn^{\prime }}^{ab}=\delta _{nn^{\prime
}}H^{ab}$, $\mathcal{C}_{mm^{\prime }}^{ab}=\delta _{mm^{\prime }}C_{m}^{ab}$%
. Using the polar decomposition (\ref{su}), we write 
\[
V(SS^{\dagger })=\tilde{U}_{R}\text{diag}(V(d_{1}^{2}),\cdots ,V(d_{n}^{2}))%
\tilde{U}_{R}^{\dagger },\qquad V(S^{\dagger }S)=U_{L}^{\dagger }\text{diag}%
(V(d_{1}^{2}),\cdots ,V(d_{n}^{2}))U_{L}. 
\]
The potential reads 
\begin{equation}
W(M)=\text{tr}[\Lambda _{L}^{2}\mathcal{H}S^{\dagger }\mathcal{C}%
S+2V(S^{\dagger }S)]=\text{tr}[\Lambda _{L}^{2}\mathcal{C}S\mathcal{H}%
S^{\dagger }+2V(SS^{\dagger })].  \label{pot}
\end{equation}
The stationary points must satisfy 
\begin{equation}
\frac{\partial W(M)}{\partial S}=\Lambda _{L}^{2}\mathcal{H}S^{\dagger }%
\mathcal{C}+2V^{\prime }(S^{\dagger }S)S^{\dagger }=0,\qquad \frac{\partial
W(M)}{\partial S^{\dagger }}=\Lambda _{L}^{2}\mathcal{C}S\mathcal{H}%
+2V(SS^{\dagger })S=0.  \label{sta}
\end{equation}
We may assume that $S$ is non-singular at the minimum. Indeed, it is not
difficult to prove, following the example treated before, that, if the free
parameters contained in $W_{2}$ satisfy suitable inequalities, the singular
configurations can be stationary points, but not minima.

Defining 
\[
H_{\Delta }=U_{L}\mathcal{H}U_{L}^{\dagger },\qquad C_{\Delta }=\tilde{U}%
_{R}^{\dagger }\mathcal{C}\tilde{U}_{R}, 
\]
equations (\ref{sta}) become 
\[
-\Lambda _{L}^{2}H_{\Delta }DC_{\Delta }=-\Lambda _{L}^{2}C_{\Delta
}DH_{\Delta }=2V^{\prime }(D^{2})D=\text{diagonal}. 
\]
We see that the matrices $\tilde{H}_{D}=\sqrt{D}H_{\Delta }\sqrt{D}$ and $%
\tilde{C}_{D}=\sqrt{D}C_{\Delta }\sqrt{D}$ are Hermitian and commute with
each other, so they can be simultaneously diagonalized with a unitary
transformation. Moreover, their product $\tilde{H}_{D}\tilde{C}_{D}$ is
itself diagonal. This means that both $\tilde{H}_{D}$ and $\tilde{C}_{D}$
are already diagonal. In turn, also $H_{\Delta }$ and $C_{\Delta }$ are
diagonal, so $U_{L}$ and $\tilde{U}_{R}$ must be matrices that diagonalize $%
\mathcal{H}$ and $\mathcal{C}$, respectively. The most general such matrices
are 
\begin{equation}
U_{L}=\left( 
\begin{tabular}{cc}
$U_{L}^\prime$ & $0$ \\ 
$0$ & $U_{L}^\prime$%
\end{tabular}
\right) U_{2},\qquad \tilde{U}_{R}=\left( 
\begin{tabular}{cc}
$\tilde{U}_{Ru}$ & $0$ \\ 
$0$ & $\tilde{U}_{Rd}$%
\end{tabular}
\right) ,  \label{matri}
\end{equation}
where $U_{L}^\prime\in U(3)$ and $\tilde{U}_{Ru}$, $\tilde{U}_{Rd}\in \tilde{%
U}_{\ell}(3)$ are unitary matrices that rotate the generations, but are
inert on the $SU(2)_{R}$- and $SU(2)_{L}$-indices $m$ and $n$, while $%
U_{2}\in SU(2)$ acts on the indices $m$, $n$, but is inert on the
generations. The reason why $U_{L}$ has this factor $U_{2}$ is that $%
\mathcal{H}$ has two coinciding diagonal blocks, which can be freely
rotated. We could factor out the unitary diagonal matrices that multiply $%
U_{L}^\prime$ to the left, as we do for the unitary diagonal matrices that
multiply $\tilde{U}_{Ru}$ and $\tilde{U}_{Rd}$ to the right, but we do not
need to.

We conclude that the non-singular stationary points have the form 
\begin{equation}
S_{\text{min}}=\left( 
\begin{tabular}{cc}
$\tilde{U}_{Ru}$ & $0$ \\ 
$0$ & $\tilde{U}_{Rd}$%
\end{tabular}
\right) D\left( 
\begin{tabular}{cc}
$U_{L}^\prime$ & $0$ \\ 
$0$ & $U_{L}^\prime$%
\end{tabular}
\right) U_{2}.  \label{simin}
\end{equation}
Arguing as before, these points are also global minima in the scalar sector
and local minima in the full $M$-space.

Now, observe that the kinetic and Yukawa terms of the action are invariant
under $G_{S}\equiv U(3)_{L}\times U(3)_{Ru}\times U(3)_{Rd}$, if the
auxiliary fields are transformed appropriately. The leading-order correction 
$\mathcal{V}(M)$ to the potential is also invariant under $G_{S}$, while the
tree-level potential $W_{2}(M)$ breaks $G_{S}$ explicitly. The $G_{S}$- and $%
SU(2)_{L}\times U(1)_{Y}$-transformations allow us to turn the minimum (\ref
{simin}) into the diagonal form $S_{\text{min}}=D$, which preserves $%
U(1)_{Q} $. Once we have done this, the diagonal entries of $D$ are the
quark masses. However, we discover that the CKM\ matrix is trivial, namely
there is no mixing among generations. Thus, our assumption (\ref{asso}) is
phenomenologically too restrictive.

In the special case 
\begin{equation}
C_{m}^{abcd}=c_{m}\delta ^{ac}\delta ^{bd},  \label{moda}
\end{equation}
the theory is completely invariant under the global symmetry $G_{S}$, which
is also preserved by renormalization. The minimum of the effective potential
does break this symmetry (because it is diagonal in the space of
generations, but not proportional to the identity). However, with the choice
(\ref{moda}) the model predicts only two different quark masses, since $%
W_{2} $ contains only two free parameters, $c_{t}$ and $c_{b}$.

The results obtained in this example generalize immediately to an arbitrary
number of generations: with a choice like (\ref{asso}) the minimum can
always be put into a diagonal form, with no mixing among generations.

A source of mixing among generations is provided by the matrix $B_{0}$,
which was taken to be proportional to the identity in this section. Now we
show that there is enough room for a non-trivial CKM matrix \textit{even if}
we still assume $B_{0}=1$. Indeed, it is sufficient to take a less symmetric
tree-level potential $W_{2}(M)$.

\paragraph{Emergence of the CKM\ matrix and mixing among generations}

Now we show that the emergence of the CKM\ matrix can be explained taking 
\begin{equation}
C_{m}^{abcd}=H_{m}^{bd}C_{m}^{ca},  \label{asso2}
\end{equation}
where $H_{m}$ and $C_{m}$ are again Hermitian matrices. We still assume $%
B_{0}=B_{1}=1$. Defining the matrices $\mathcal{H}_{1\hspace{0.01in}%
nn^{\prime }}^{ab}=\delta _{nn^{\prime }}H_{1}^{ab}$, $\mathcal{H}_{2\hspace{%
0.01in}nn^{\prime }}^{ab}=\delta _{nn^{\prime }}H_{2}^{ab}$, $\mathcal{C}_{1%
\hspace{0.01in}mm^{\prime }}^{ab}=\delta _{m1}\delta _{m^{\prime
}1}C_{1}^{ab}$, $\mathcal{C}_{2\hspace{0.01in}mm^{\prime }}^{ab}=\delta
_{m2}\delta _{m^{\prime }2}C_{2}^{ab}$, now the potential reads 
\begin{equation}
W(M)=\text{tr}[\Lambda _{L}^{2}S\mathcal{H}_{1}S^{\dagger }\mathcal{C}%
_{1}+\Lambda _{L}^{2}S\mathcal{H}_{2}S^{\dagger }\mathcal{C}%
_{2}+2V(S^{\dagger }S)].  \label{pot2}
\end{equation}
The stationary points are the solutions of 
\[
\tilde{H}_{1D}\tilde{C}_{1D}+\tilde{H}_{2D}\tilde{C}_{2D}=\tilde{C}_{1D}%
\tilde{H}_{1D}+\tilde{C}_{2D}\tilde{H}_{2D}=-\frac{2}{\Lambda _{L}^{2}}%
V^{\prime }(D^{2})D^{2}=\text{diagonal}, 
\]
where 
\[
\tilde{H}_{mD}=\sqrt{D}U_{L}\mathcal{H}_{m}U_{L}^{\dagger }\sqrt{D},\qquad 
\tilde{C}_{mD}=\sqrt{D}\tilde{U}_{R}^{\dagger }\mathcal{C}_{m}\tilde{U}_{R}%
\sqrt{D},\qquad m=1,2. 
\]

If we search for a solution of the form 
\begin{equation}
S_{\text{min}}=\left( 
\begin{tabular}{cc}
$\tilde{U}_{Ru}$ & $0$ \\ 
$0$ & $\tilde{U}_{Rd}$%
\end{tabular}
\right) D\left( 
\begin{tabular}{cc}
$U_{Lu}$ & $0$ \\ 
$0$ & $U_{Ld}$%
\end{tabular}
\right) U_{2}  \label{smin}
\end{equation}
and argue as before, we find that $U_{Lu}$, $U_{Ld}\in U(3)$, $\tilde{U}%
_{Ru} $, $\tilde{U}_{Rd}\in \tilde{U}_{\ell}(3)$ must be matrices that
diagonalize $H_{1}$, $H_{2} $, $C_{1}$, $C_{2}$, respectively.

At this point we can proceed as usual: the invariance of the rest of the
action under phase transformations and $SU(2)_{L}\times U(1)_{Y}\times G_{S}$
allows us to turn the minimum into the form 
\begin{equation}
S_{\text{min}}^{\prime }=\left( 
\begin{tabular}{cc}
$1$ & $0$ \\ 
$0$ & $C_{KM}$%
\end{tabular}
\right) D,  \label{spmin}
\end{equation}
which preserves $U(1)_{Q}$, where $C_{KM}$ is the CKM matrix. This
stationary point can describe the properties of the Standard Model at low
energies.

We have only proved that (\ref{spmin}) belongs to the set of extremal points
of the potential. Strictly speaking, there could be other extrema that are
not block diagonal, and therefore spontaneously break also charge
conservation.

If we take the most general potential (\ref{mosto}) every minimum that
preserves $U(1)_{Q}$ can be cast into the form (\ref{spmin}). Indeed, $%
U(1)_{Q}$-conservation means that the charged $S$-entries, which are $%
S_{12}^{ab}$ and $S_{21}^{ab}$, vanish, therefore the minimum is
block-diagonal. Then it can be turned to the form (\ref{spmin}) arguing as
before, namely using invariance under phase transformations and $%
SU(2)_{L}\times U(1)_{Y}\times G_{S}$.

Finally, let us comment about the case $B_{0}\neq 1$. If $B_{0}$ is not
diagonal it can be diagonalized using $SU(3)_{L}\times SU(3)_{Ru}\times
SU(3)_{Rd}$. Then we cannot use such transformations to turn (\ref{smin})
into the form (\ref{spmin}). We can only simplify (\ref{smin}) by means of
(eight) phase transformations. So, the Lorentz violation predicts more
mixing among generations besides the CKM matrix. It also predicts mixing
among leptons. If leptons have a non-diagonal matrix $B_{0\ell }$, we can
use the freedom we have to diagonalize it, but then the lepton mass matrix
remains non-diagonal.

If both $B_{0}$ and $B_{1}$ are different from the identity, we can
diagonalize only one of them for each particle.

\section{CPT\ violating local minima}

\setcounter{equation}{0}

In this section we want to show that the effective potential may also give
non-trivial expectation values to the vector and tensor fields $H_{\mu
},K_{\mu },L_{\mu \nu }$. For simplicity, we assume $B_{0}=B_{1}=1$ and
concentrate on the vector $H_{\mu }$.

The most general tree-level potential with one generation is 
\[
W_{2}(M)=\Lambda _{L}^{2}\left( c_{1}\text{tr}[H_{0}]^{2}+c_{2}\text{tr}[%
H_{i}]^{2}+c_{3}\text{tr}[H_{0}^{2}]+c_{4}\text{tr}[H_{i}^{2}]\right) , 
\]
where $c_{1\text{-}4}$ are constants. After simple manipulations, the
one-loop correction can be expressed in the form 
\[
\mathcal{V}(M)=-N_{c}\int\frac{\text{\textrm{d}}^{4}p}{(2\pi )^{4}}\left[
\ln \det (A+\sigma _{i}B_{i})-\frac{4\text{tr}[H_{i}^{2}](\bar{p}^{\prime
})^{2}}{3(p^{\prime \hspace{0.01in}2})^{2}}\right] , 
\]
where 
\[
A=1+\frac{1}{(p^{\prime })^{2}}(ip_{4}H_{0}-p_{i}^{\prime }H_{i}),\qquad
B_{i}=\frac{1}{(p^{\prime })^{2}}(-ip_{4}H_{i}+p_{i}^{\prime
}H_{0}-ip_{j}^{\prime }H_{k}\varepsilon _{ijk}). 
\]
However, since $H_{0}$ and $H_{i}$ are 2$\times $2 matrices, it is still
difficult to evaluate $\mathcal{V}(M)$ explicitly. If we restrict to the
case of a single fermion we can perform the calculation to the end. We find 
\begin{eqnarray*}
W_{2}(M) &=&\Lambda _{L}^{2}(c_{1}^{\prime }H_{0}^{2}+c_{2}^{\prime
}H_{i}^{2}), \\
\mathcal{V}(M) &=&\frac{N_{c}\Lambda _{L}^{4}}{7560h\pi ^{2}}\left[
630h^{3}\ln (v^{2}+1)-v^{3}\left(
140v^{6}+360v^{4}-630hv^{3}+252v^{2}-945hv+1260h^{2}\right) \right] ,
\end{eqnarray*}
with 
\[
v=\frac{2^{1/3}\Delta ^{2/3}-2\cdot 3^{1/3}}{6^{2/3}\Delta ^{1/3}},\qquad
\Delta =\sqrt{12+81h^{2}}+9h,\qquad h=\sqrt{\frac{H_{i}^{2}}{\Lambda _{L}^{2}%
}}. 
\]
The one-loop correction $\mathcal{V}(M)$ does not depend on $H_{0}$, so to
have a minimum we must assume $c_{1}^{\prime }>0$. As a function of $h$, $%
\mathcal{V}(M)$ is monotonic and convex, and $\mathcal{V}(M)=\mathcal{O}%
(H^{4})$ in a neighborhood of the origin. Thus, we have two phases:

\noindent 1) the unbroken phase has $c_{1}^{\prime }>0$, $c_{2}^{\prime }>0$;

\noindent 2) the broken phase has 
\[
c_{1}^{\prime }>0,\qquad c_{2}^{\prime }<0. 
\]
where $H$ has a non-trivial expectation value. Here the minimum of the
effective potential spontaneously breaks invariance under boosts, rotations
and CPT.

In the simple example just studied, the potential $\mathcal{V}(M)$ does not
depend on $H_{0}$. The reason is that $H_{0}$ can be reabsorbed with an
imaginary translation of $p_{4}$. Observe that $H_{i}$ cannot be reabsorbed
away. Indeed, although the integrand depends only on the sum $p_{i}^{\prime
}+H_{i}$, we cannot translate $p_{i}^{\prime }$, because the integral is in $%
p_{i}$, not in $p_{i}^{\prime }$. On the other hand, only one $p_{4}$
translation is available, so we expect that with more fermions, where $H_{0}$
is a matrix, there exist broken phases where $H_{i}=0$ but some entries of
the $H_{0}$-matrix get non-trivial expectation values. In such phases CPT\
and boosts are broken, but rotations are preserved.

%

\section{Low-energy effective action}

\setcounter{equation}{0}

In this section we study the low-energy effective action in the Lorentz
phase. We work at the leading order of the $1/N_{c}$ expansion, at $%
B_{0}=B_{1}=1$, and focus on the third generation. As usual, we first turn
the gauge-field interactions off and turn them back on at a second stage. We
study the spectrum of composite bosons, derive a number predictions and show
that the model is compatible with the experimental data. For the moment we
can concentrate on the scalar sector.

To keep the presentation readable, at first we assume not only invariance
under $SU(2)_{L}\times U(1)_{Y}\times U(1)_{B}$, but also the axial symmetry 
$U(1)_{A}$. With this assumption, however, the low-energy model is ruled out
by experimental data. It is straightforward to relax the assumption of axial
symmetry at a second stage. We show that once $U(1)_{A}$ is explicitly
broken full compatibility with data is achieved.

We refer to section 5 for the notation. The total four-fermion Lagrangian is 
$\mathcal{L}_{\text{tot}}=\mathcal{L}_{q}+\mathcal{L}_{\ell }$, where the
quark- and lepton-contributions are 
\begin{eqnarray}
\mathcal{L}_{q} &=&\bar{\Psi}\left( i\Gamma ^{0}\mathds{1}\partial _{t}+i%
\bar{\Gamma}\cdot {\bar{\partial}}\left( 1-\frac{{\bar{\partial}}^{2}}{%
\Lambda _{L}^{2}}\right) -M\right) \Psi -\Lambda _{L}^{2}\text{tr}[\tau \tau
^{\dagger }C]  \label{ono} \\
\mathcal{L}_{\ell } &=&\mathcal{L}_{\ell \text{kin}}-\sum_{ab}\left(
y^{ab}\tau _{2n}\bar{\ell}_{R}^{a}L_{n}^{b}+\bar{y}^{ba}\bar{L}_{n}^{a}\ell
_{R}^{b}\bar{\tau}_{2n}\right) ,  \label{odo}
\end{eqnarray}
$y^{ab}$ being constants, while $\Psi =((t_{L},b_{L}),(t_{R},b_{R}))$. The
form of $\mathcal{L}_{\ell }$ is justified as follows.

Since we are working in the leading order of the $1/N_{c}$ expansion, we
have to calculate one-loop diagrams with circulating quarks. Thus, we can
focus on four-fermion vertices that contain two quarks $q$ and two leptons $%
\ell $, or four quarks, and ignore the vertices that contain four leptons.
Introducing auxiliary scalar fields $\tau $ and $\sigma $, as usual, we get
Yukawa and potential terms of the form 
\[
-\tau qq-\sigma \ell \ell -\frac{a}{2}\tau ^{2}-b\tau \sigma -\frac{c}{2}%
\sigma ^{2}. 
\]
The leading-order correction $\mathcal{V}$ to the potential depends only on $%
\tau $, so the effective potential has the form 
\[
\mathcal{W}(\tau ,\sigma )=\frac{a}{2}\tau ^{2}+b\tau \sigma +\frac{c}{2}%
\sigma ^{2}+\mathcal{V}(\tau ). 
\]
Its extrema can also be found replacing $\sigma $ with the solution $\sigma
=-b\tau /c$ of its field equation, namely working with $\mathcal{W}(\tau
,-b\tau /c)$. Therefore, we do not need to multiply the lepton bilinears $%
\ell \ell $ by independent auxiliary scalars $\sigma $. We can just multiply
them by entries of $\tau $ and free parameters. Because of the symmetries we
have assumed, (\ref{odo}) is the only form that is allowed. Moreover, using
the polar decomposition on $y^{ab}$ and performing unitary transformations
on $L^{a}$ and $\ell _{R}^{a}$, we can diagonalize the matrices $y^{ab}$.
Thus, from now on we take $y^{ab}=\delta ^{ab}$diag$(y^{a})$, with $y^{a}$
real.

We expand around the minimum (\ref{desi}), writing $\tau =\tau _{0}+\eta $.
We first recall the leading contributions to the quadratic effective action $%
\Gamma _{2}$ \cite{noH}, namely 
\begin{equation}
\Gamma _{2}=-N_{c}\sum_{ij}\eta _{ij}\left( \partial ^{2}+2m_{j}^{2}\right)
f_{ij}\bar{\eta}_{ij}-N_{c}\sum_{ij}m_{i}m_{j}f_{ij}(\eta _{ij}\eta _{ji}+%
\bar{\eta}_{ij}\bar{\eta}_{ji})  \label{gamma2}
\end{equation}
(the constants $f_{ij}$ being defined in Appendix B and the integration over
spacetime being understood), which gives the following propagating fields: $%
i $) two neutral massive scalars $\varphi _{1,2}$ and a charged massive
scalar $\varphi $, 
\[
\varphi _{1}=\sqrt{2N_{c}f_{tt}}\,\mathrm{Re}\,\eta _{tt},\qquad \varphi
_{2}=\sqrt{2N_{c}f_{bb}}\,\mathrm{Re}\,\eta _{bb},\qquad \varphi =\sqrt{%
\frac{N_{c}f_{tb}}{m_{t}^{2}+m_{b}^{2}}}\left( m_{b}\eta _{tb}+m_{t}\bar{\eta%
}_{bt}\right) , 
\]
with squared masses 
\[
m_{1}^{2}=4m_{t}^{2},\qquad m_{2}^{2}=4m_{b}^{2},\qquad m^{2}=2\left(
m_{t}^{2}+m_{b}^{2}\right) , 
\]
respectively; $ii$) the Goldstone bosons associated with the spontaneously
broken generators of $SU(2)_{L}\times U(1)_{Y}$, which are

\[
\phi ^{+}=i\sqrt{\frac{N_{c}}{2f_{W}}}f_{tb}(m_{t}\eta _{tb}-m_{b}\bar{\eta}%
_{bt}),\qquad \phi ^{0}=\sqrt{\frac{N_{c}}{f_{Z}}}(m_{b}f_{bb}\,\mathrm{Im}%
\,\eta _{bb}-m_{t}f_{tt}\,\mathrm{Im}\,\eta _{tt}), 
\]
and $\phi ^{-}=\bar{\phi}^{+}$, where 
\[
f_{W}=\frac{f_{tb}}{2}(m_{t}^{2}+m_{b}^{2}),\qquad f_{Z}=\frac{1}{2}\left(
m_{t}^{2}f_{tt}+m_{b}^{2}f_{bb}\right) ; 
\]
$iii$) a Goldstone boson 
\[
\tilde{\phi}^{0}=\sqrt{\frac{N_{c}f_{bb}f_{tt}}{f_{Z}}}\left( m_{b}\,\mathrm{%
Im}\,\eta _{tt}+m_{t}\,\mathrm{Im}\,\eta _{bb}\right) , 
\]
associated with the broken axial symmetry.

When gauge interactions are switched back on, the Goldstone bosons $\phi
^{\pm ,0}$ are ``eaten'' by the gauge fields. Then the gauge fields acquire
squared masses 
\begin{equation}
m_{W}^{2}=N_{c}g^{2}f_{W},\qquad m_{Z}^{2}=N_{c}\tilde{g}^{2}f_{Z}.
\label{mwmz}
\end{equation}

Including the covariant derivatives for $U(1)_{Q}$ the quadratic effective
action $\Gamma _{2}$ becomes 
\begin{eqnarray*}
\Gamma _{2} &=&\frac{1}{2}\sum_{i=1}^{2}\left[ (\partial _{\mu }\varphi
_{i})(\partial ^{\mu }\varphi _{i})-m_{i}^{2}\varphi _{i}^{2}\right]
+(\partial _{\mu }\bar{\varphi}-ieA_{\mu }\bar{\varphi})(\partial ^{\mu
}\varphi +ieA^{\mu }\varphi )-m^{2}\bar{\varphi}\varphi +\frac{1}{2}\partial
_{\mu }\tilde{\phi}^{0}\partial ^{\mu }\tilde{\phi}^{0} \\
&&+(\partial _{\mu }\phi ^{+}-m_{W}W_{\mu }^{+})(\partial ^{\mu }\phi
^{-}-m_{W}W^{\mu -})+\frac{1}{2}(\partial _{\mu }\phi ^{0}-m_{Z}Z_{\mu
})(\partial ^{\mu }\phi ^{0}-m_{Z}Z^{\mu })
\end{eqnarray*}
and it is invariant under the linearized gauge transformations 
\begin{equation}
\delta W_{\mu }^{\pm }=\partial _{\mu }C^{\pm },\qquad \delta Z_{\mu
}=\partial _{\mu }C^{0},\qquad \delta \phi ^{\pm }=m_{W}C^{\pm },\qquad
\delta \phi ^{0}=m_{Z}C^{0}.  \label{ga}
\end{equation}

Now we calculate the three-leg and four-leg terms $\Gamma _{3}$ and $\Gamma
_{4}$ of the effective action. We focus on the terms proportional to factors
of the form $\ln (\Lambda _{L}^{2}/m^{2})$, where $m$ is a function of the
masses, because they are numerically more important, in our approximation.
We find (again, refer to Appendix B for the notation) 
\begin{equation}
\Gamma _{3}+\Gamma _{4}=-2N_{c}\sum_{ijk}m_{i}f_{ijk}(\eta _{ij}\bar{\eta}%
_{kj}\eta _{ki}+\bar{\eta}_{ij}\eta _{kj}\bar{\eta}_{ki})-N_{c}%
\sum_{ijkl}f_{ijkl}\eta _{ij}\bar{\eta}_{kj}\eta _{kl}\bar{\eta}_{il}.
\label{g3g4}
\end{equation}

Writing (\ref{gamma2}) and (\ref{g3g4}) we have omitted some terms that are
numerically negligible. Basically, they do not contain the enhancing factor $%
\sim \ln \Lambda _{L}^{2}$. Examples of such terms are 
\begin{equation}
\frac{N_{c}}{24\pi ^{2}}(\partial _{\mu }\,\mathrm{Re}\,\eta _{tt})(\partial
^{\mu }\,\mathrm{Re}\,\eta _{tt}),\qquad \frac{N_{c}m_{t}}{3(4\pi )^{2}}\eta
_{tt}^{3},\qquad \frac{N_{c}m_{b}}{3(4\pi )^{2}}\eta _{bb}^{3},\qquad \frac{%
2N_{c}m_{b}}{(4\pi )^{2}}\eta _{tt}\eta _{tb}\eta _{bt},  \label{r3}
\end{equation}
(using $m_{t}\gg m_{b}$). We can compare them with the smallest cubic term
in (\ref{g3g4}), which is 
\begin{equation}
-\frac{2N_{c}}{(4\pi )^{2}}m_{b}(\eta _{bj}\bar{\eta}_{kj}\eta _{kb}+\bar{%
\eta}_{bj}\eta _{kj}\bar{\eta}_{kb})\ln \frac{\Lambda _{L}^{2}}{m_{t}^{2}}
\label{r4}
\end{equation}
Numerically, with $\Lambda _{L}=10^{14}$GeV and using $m_{t}=171.2$GeV, $%
m_{b}=4.2$GeV, we find that the coefficient of the second term of (\ref{r3})
is about 13\% of the coefficient of (\ref{r4}). All other terms of type (\ref
{r3}) are suppressed by a factor $1/\ln (\Lambda _{L}^{2}/m_{t}^{2})$, which
is a 2\%. In any case, these contributions are below our errors. Moreover,
since $\ln (\Lambda _{L}^{2}/m_{t}^{2})$ and $\ln (\Lambda
_{L}^{2}/m_{b}^{2})$ differ only by a 14\%, we can also neglect their
difference and replace $m_{b}$ with $m_{t}$ inside the logarithms. Finally,
the recurring factor 
\[
\mathcal{N}\equiv \frac{N_{c}}{(4\pi )^{2}}\ln \frac{\Lambda _{L}^{2}}{%
m_{t}^{2}} 
\]
can be approximated to one up to a negligible 3\%. However, we continue to
write it down explicitly, to keep track of the $\Lambda _{L}$-dependence.

Collecting $\Gamma _{2}$, $\Gamma _{3}$ and $\Gamma _{4}$, we get the
low-energy scalar effective action 
\begin{equation}
\Gamma \sim \mathcal{N}\text{tr}[\partial _{\mu }\tau \partial ^{\mu }\tau
^{\dagger }+2\tau _{0}^{2}\tau \tau ^{\dagger }-\tau \tau ^{\dagger }\tau
\tau ^{\dagger }],  \label{lowe}
\end{equation}
which is a type II\ two Higgs doublet model (2HDM), namely a model with two
Higgs doublets, where one doublet couples only to top quarks, while the
other doublet couples only to bottom quarks and leptons.

Because of the assumed axial symmetry $U(1)_{A}$, the scenario explored so
far is ruled out by data. Indeed, it predicts very light neutral Higgs
bosons, such as the field $\varphi _{2}$ of mass $\sim 2m_{b}$ and the
massless $U(1)_{A}$ Goldstone boson. These fields violate the present
experimental lower bound on the mass of neutral Higgs bosons, which is
114GeV \cite{114}. This bound, established through the process $Z\rightarrow
Zh\rightarrow Z\bar{b}b$, applies to our model. Indeed, take for example the
field $\varphi _{2}$ as the Higgs boson $h$. It is easy to check that
although the vertex $ZZh$ is suppressed by a factor $m_{b}/m_{t}$, the
Yukawa coupling $h\bar{b}b$ is enhanced by the reciprocal factor $%
m_{t}/m_{b} $, so the process $Z\rightarrow Zh\rightarrow Z\bar{b}b$ is not
suppressed with respect to one predicted by the minimal Standard Model.

Compatibility with data can be obtained breaking $U(1)_{A}$ explicitly.

\paragraph{Low-energy model compatible with data}

It is easy to see that, because of $SU(2)_{L}\times U(1)_{Y}$ invariance,
the $U(1)_{A}$ symmetry can be explicitly broken in a unique way by
four-fermion vertices. Indeed, only one term can be added to the tree-level
potential $W_{2}$, namely 
\begin{equation}
\Delta W_{2}=\tilde{m}_{12}^{2}\text{tr}\left[ \tau \epsilon \tau
^{T}\epsilon \right] +\tilde{m}_{12}^{*2}\text{tr}\left[ \tau ^{\dagger
}\epsilon \tau ^{*}\epsilon \right] \text{,}  \label{Acorr}
\end{equation}
where $^{T}$ denotes transposition, $\epsilon _{tt}=\epsilon _{bb}=0$, $%
\epsilon _{tb}=-\epsilon _{bt}=1$, and $\tilde{m}_{12}$ is a complex
constant. The one-loop correction $\mathcal{V}$ is unaffected, therefore
still $U(1)_{A}$-symmetric. The term (\ref{Acorr}) displaces the minimum and
changes the mass spectrum.

For simplicity, we take $\tilde{m}_{12}$ real. To bring the displaced
minimum back to the form (\ref{desi}), we also modify the term 
\[
2\tau _{0}^{2}\tau \tau ^{\dagger } 
\]
of (\ref{lowe}) replacing $\tau _{0}^{2}$ with a different diagonal matrix.
With our approximations we find the low-energy type II 2HDM\ Lagrangian 
\begin{equation}
\Gamma =\ \mathcal{N}\text{tr}\left[ \partial _{\mu }\tau \partial ^{\mu
}\tau ^{\dagger }+2\tau _{0}^{2}\tau \tau ^{\dagger }-\frac{%
m_{12}^{2}m_{t}m_{b}}{2(m_{t}^{2}+m_{b}^{2})}\left( \tau \epsilon \tau
^{T}\epsilon +\tau ^{\dagger }\epsilon \tau ^{*}\epsilon -2\epsilon \tau
_{0}\epsilon \tau _{0}^{-1}\tau \tau ^{\dagger }\right) -\tau \tau ^{\dagger
}\tau \tau ^{\dagger }\right] ,  \label{other}
\end{equation}
Expanding $\tau $ as $\tau _{0}+\eta $ we can first check that the minimum
is still $\tau _{0}$, and then work out the new spectrum. We find that,
using $m_{b}\ll m_{t}$,

\noindent $i$) the three Goldstone bosons $\phi ^{\pm ,0}$ associated with
the $SU(2)_{L}\times U(1)_{Y}$ symmetry are unaffected,

\noindent $ii$) the mass of the charged composite Higgs boson $\varphi $
becomes 
\[
m_{\varphi }\sim \sqrt{2m_{t}^{2}+m_{12}^{2}}\hspace{0.01in}, 
\]

\noindent $iii$) assuming also $m_{b}\ll m_{12}$, the masses of the neutral
Higgs bosons $\varphi_1$ and $\varphi_2$ become 
\[
m_{12},\qquad 2m_{t}\text{\hspace{0.01in}}, 
\]
which is which depending on whether $m_{12}>2m_{t}$ or $m_{12}<2m_{t}$,

\noindent $iv$) the field $\tilde{\phi}^{0}$ acquires a mass equal to $%
m_{12} $,

\noindent $v$) the neutral fields $(\varphi_1,\varphi_2)$ are rotated by an
angle $\alpha$, while all other fields preserve the expressions they had
before.

Since four fermion vertices are multiplied by $1/\Lambda _{L}^{2}$, the
tree-level potential terms, such as (\ref{Acorr}), are proportional to $%
\Lambda _{L}^{2}$, which means that $m_{12}$ is large. For $m_{12}$
sufficiently large the masses of all particles become compatible
with data. Taking into account of our errors ($\pm $50\%), even a Higgs mass
predicted to be around $2m_{t}$ could in the end be more close to $m_{t}$,
which is contained in the present mass range for Higgs boson. 

Moreover, because of $i$) the gauge-boson masses are unaffected, and
formulas (\ref{mwmz}) still hold. The Fermi constant and the parameter $\rho 
$ are given by the relations \cite{noH} 
\begin{equation}
\frac{1}{G_{F}}=\frac{N_{c}m_{t}^{2}}{4\pi ^{2}\sqrt{2}}\ln \frac{\Lambda
_{L}^{2}}{m_{t}^{2}},\qquad \rho =\frac{\tilde{g}^{2}m_{W}^{2}}{%
g^{2}m_{Z}^{2}}\sim 1.  \label{fermi}
\end{equation}

Formulas (\ref{fermi}) provide two important checks of our model. The
Standard Model provides no analogue of the first formula. At $\Lambda
_{L}=10^{14}$GeV the first prediction turns out to be very precise. As far
as $\rho $ is concerned, the Standard Model predicts $\rho =1$ up to
radiative corrections, which matches experimental data very well. Our
approach is consistent with this, but cannot be equally precise, because our
theoretical errors are large.

So far, we have focused on the scalar sector and ignored the fields $H_{\mu
} $, $K_{\mu }$ and $L_{\mu \nu }$. It is easy to prove, computing their
two-point functions in the low-energy limit, that such fields do become
propagating at some point. Moreover, the dominant contributions to their
kinetic terms, namely the contributions proportional to $\ln \Lambda
_{L}^{2} $, are Lorentz invariant. Thus, our model also predicts composite
vectors and tensors at low energies. Nevertheless, it is unable to predict
their masses, whose values can be changed adding quadratic terms
proportional to $H^{2}$, $K^{2}$ and $L^{2}$ to the tree-level potential $%
W_{2}(M)$, multiplied by 
coefficients proportional to $\Lambda_{L}^{2}$. The basic reason is that in the Lorentz phase $H_{\mu }$, $K_{\mu
}$ and $L_{\mu \nu }$ have trivial gap equations. 
Thus, we are free to assume that the masses of these fields are sufficiently large, in which case
this subsector of our model is also compatible with data.

\paragraph{The limit $m_{12}\rightarrow \infty $}

The limit $m_{12}\rightarrow \infty $ is particularly interesting, because
it gives the usual one-doublet model. The coefficient of $m_{12}^{2}$ in (%
\ref{other}) must vanish in the limit, which requires 
\begin{equation}
\tau =u\left( 
\begin{tabular}{cc}
$H_{2}$ & $-H_{1}$ \\ 
$\kappa \bar{H}_{1}$ & $\kappa \bar{H}_{2}$%
\end{tabular}
\right) ,\qquad \kappa =\frac{m_{b}}{m_{t}},\qquad u^{-2}=(1+\kappa ^{2})%
\mathcal{N}.  \label{hig}
\end{equation}
Then we find a particular case of the usual Higgs Lagrangian, namely (using
again $m_{t}\gg m_{b}$) 
\begin{equation}
\Gamma _{H}=\partial _{\mu }H^{\dagger }\partial ^{\mu }H-V(H),\qquad
V(H)=2m_{t}^{2}H^{\dagger }H-u^{2}(H^{\dagger }H)^{2}.  \label{onedou}
\end{equation}
From this formula we can read: $i$) the Higgs vacuum expectation value ($%
|H|_{\text{min}}=v/\sqrt{2}$), which is (with $\Lambda _{L}=10^{14}$GeV) 
\[
v=\frac{m_{t}}{u}\sqrt{2}\sim 247\text{GeV}, 
\]
$ii$) the constant 
\[
\lambda =u^{2}\sim 1\text{,} 
\]
and, consequently, $iii$) the Higgs-boson mass, which is $2m_{t}$.

The Yukawa couplings are automatically correct. We have 
\begin{eqnarray}
\mathcal{L}_{\text{Yukawa}} &=&-\frac{m_{t}}{v}\sqrt{2}\left( \bar{t}_{R}%
\tilde{H}Q_{L}+\bar{Q}_{L}t_{R}\tilde{H}^{\dagger }\right) -\frac{m_{b}}{v}%
\sqrt{2}\left( H^{\dagger }\bar{b}_{R}Q_{L}+\bar{Q}_{L}b_{R}H\right) 
\nonumber \\
&&-\frac{\sqrt{2}}{v}\sum_{a=1}^{3}m_{\ell }^{a}\left( H^{\dagger }\bar{\ell}%
_{R}^{a}L^{a}+\bar{L}^{a}\ell _{R}^{a}H\right) ,  \label{Lyukawa}
\end{eqnarray}
where $\tilde{H}_{n}=\varepsilon _{nq}H_{q}$ and $m_{\ell }^{a}=m_{b}y^{a}$.
The lepton mass terms do not give new predictions, but just determine the
Yukawa parameters $y^{a}$.

The one-doublet model (\ref{onedou}) was already considered in \cite{noH},
but not fully justified there (it was presented as a subsector of the model
with $m_{12}=0$). The limit $m_{12}\rightarrow \infty $ provides the missing
justification for (\ref{onedou}).

\section{Neutrino masses and neutrino oscillations}

\setcounter{equation}{0}

Among the compatibility checks we can make, we mention neutrino
oscillations. In this section we show that the minimal versions of our
models cannot give masses to neutrinos and discuss alternative ways to
explain neutrino oscillations.

First, we prove that the vertex 
\begin{equation}
\frac{1}{\Lambda _{L}}(LH)^{2}=\frac{1}{\Lambda _{L}}\sum_{a,b=1}^{3}Y_{ab}\
(L_{m}^{\alpha a}\varepsilon _{\alpha \beta }L_{p}^{\beta b})\ \varepsilon
_{mn}H_{n}\varepsilon _{pq}H_{q}+\text{H.c.,}  \label{lh2}
\end{equation}
which gives Majorana masses to the neutrinos when $H$ is replaced by its
expectation value, cannot be generated.

The vertex (\ref{lh2}) breaks the conservation of $B-L$ by two units.
However, the vacuum we are considering does not break $B-L$ spontaneously.
Moreover, the global $B-L$ symmetry is anomaly-free in our model. The reason
is that it is anomaly-free in the minimal Standard Model \cite{weinbergbook}%
, and anomalies are unaffected by the Lorentz violation (see \cite{lvsm}).
Finally, the $B-L$ symmetry cannot be explicitly violated in the model (\ref
{noH}), because

\medskip \textbf{Theorem 1} \textit{all CPT invariant four-fermion vertices
constructed with the fields of the minimal Standard Model preserve }$B-L$.
\medskip

This theorem is a simple generalization of a well-known property stating the
same conclusion about Lorentz invariant four-fermion vertices \cite
{weinberg4f}. We stress here that it is not necessary to assume Lorentz
symmetry, because CPT\ is sufficient. The theorem can be proved writing down
all four-fermion vertices that are invariant under $SU(2)_{L}\times U(1)_{Y}$
and using a property proved in ref. \cite{4f}, stating that all four-fermion
vertices of the form $\ell \ell \ell \ell $ and $\ell \ell \ell ^{*}\ell
^{*} $ are CPT invariant, and all four-fermion vertices of the form $\ell
\ell \ell \ell ^{*}$ are CPT\ violating, $\ell $ denoting a left-handed
fermion.

For the sake of completeness, we write the structures of four-fermion
vertices with non-vanishing $\Delta B=\Delta L$. They are 
\begin{equation}
LQ_{L}^{3},\qquad Q_{L}^{2}u_{R}\ell _{R},\qquad LQ_{L}u_{R}d_{R},\qquad
u_{R}^{2}d_{R}\ell _{R},  \label{verto}
\end{equation}
plus their Hermitian conjugates. They all have $|\Delta B|=|\Delta L|=1$.
Such vertices do not affect the effective potential at the leading order of
the $1/N_{c}$ expansion.

The $B-L$ symmetry could be spontaneously broken at subleading orders.
However, we are not going to explore this possibility here.

Were it present, the vertex (\ref{lh2}) could explain neutrino masses with a
scale $\Lambda _{L}$ around 10$^{14}$-10$^{15}$GeV. However, it has been
speculated \cite{glashow1,kosteleckymewes} that in Lorentz violating models
neutrino masses may not be necessary to explain neutrino oscillations. We
make some observations about this fact in the realm of our models.

In the minimal model (\ref{noH}), the energies of neutrinos with given
momentum $p$ are the eigenvalues of the matrix 
\[
\mathcal{H}=p\left( b_{\nu 1}+b_{\nu 0}\frac{p^{2}}{\Lambda _{L}^{2}}\right)
, 
\]
where $b_{\nu 1}$ and $b_{\nu 0}$ are constant Hermitian matrices. In the
simple case of two generations, the mixing probability after traveling a
distance $\ell $ is 
\[
P_{\text{mixing}}=\left( 1-\frac{(\text{tr}[\Delta \mathcal{H}\sigma _{z}%
])^{2}}{\Omega ^{2}}\right) \sin ^{2}\left( \frac{\ell \Omega }{2}\right)
,\qquad \text{where }\Omega =\sqrt{2\text{tr}[\Delta \mathcal{H}^{2}]-(%
\text{tr}[\Delta \mathcal{H}])^{2}}, 
\]
where $\Delta \mathcal{H}$ is $\mathcal{H}$ minus any contribution
proportional to the identity matrix.

If was shown in ref. \cite{kmewes2} that several existing data about
neutrino oscillations can be accounted for by the matrix $b_{\nu 1}-1$. The
values of its entries were determined to be around $10^{-17}$-$10^{-22}$,
which are compatible with our approach. A different class of massless models
(with five parameters) was considered in ref. \cite{barger}, and shown to be
unable to explain all combined data about neutrino oscillations. The models
considered in ref. \cite{barger} explore a region of parameter space that is
absent in our approach, because they contain four CPT-violating parameters
out of five. At present, the problem to construct massless Lorentz-violating
models that are globally compatible with data is still open and challenging.
We suggest that it may be considered in a fully CPT invariant framework
first.

Higher-derivative corrections do not appear to be helpful here. If we wanted
to explain neutrino oscillations using only $b_{\nu 0}$ (setting $b_{\nu
1}=1 $), we would find $b_{\nu 0}\gg 1$ by several orders of magnitude. We
expect that large $b_{\nu 0}$-values are unlikely. The matrices $b_{0}$ have
been studied in other sectors of the model, particularly quantum
electrodynamics \cite{taiuti2}, and found to be small or at most of order 1.
Thus, the effects of terms containing higher-space derivatives are expected
to be negligible for neutrino oscillations. Non-minimal versions of our
model can be considered, and certainly have the chance to account for all
data. Nevertheless, there is still hope that neutrino oscillations can be
fully accounted for by the sole matrix $b_{\nu 1}-1$ in the minimal
scalarless model.

\section{Conclusions and outlook}

\setcounter{equation}{0}

In this paper we have studied the low-energy phenomenology of renormalizable
CPT invariant Standard Model extensions that violate Lorentz symmetry at
high energies. These models include operators of higher dimensions, in
particular four-fermion vertices, and contain no elementary scalar fields.
At the leading order of the large $N_{c}$ expansion, a dynamical
symmetry-breaking mechanism gives masses to fermions and gauge bosons, and
generates composite scalars. We have studied the effective potential and the
phase diagram. A broken phase always exists. In general, it may break
boosts, rotations and CPT. We have given evidence that there exists a
Lorentz phase, described the mixing among generations and the emergence of
the CKM matrix.

The low-energy effective action in the Lorentz phase looks like a Standard
Model with one or more Higgs doublets, and possibly very heavy composite
vectors and tensors. Not all parameters are free, but some are related by
formulas induced by the high-energy model. For example, our approach gives a
formula relating the Fermi constant, the top mass and the scale of Lorentz
violation $\Lambda _{L}$. So far, our predictions are compatible with
present data, within theoretical errors.

We have considered the minimal version of our Lorentz-violating
Standard-Model extensions and made certain assumptions to simplify
calculations (such as $B_{1}=B_{0}=1$). When such assumptions are relaxed
new effects appear, such as lepton mixing and a more severe quark mixing. It
would be interesting to explore these aspects further and study the
low-energy Lagrangian with $B_{0}$ generic. Another topic for future
investigations is to explore the lowest energies where we can find remnants
of the Lorentz violation, then look for the effects that can be tested in
existing or planned experiments. It would also be interesting to explore
more general models, and include right-handed neutrinos and elementary
scalars.

\vskip 20truept \noindent {\Large \textbf{Acknowledgments}}

\vskip 10truept

D.Anselmi wishes to thank Xinmin Zhang and the Institute of High Energy
Physics of the Chinese Academy of Sciences, Beijing, for hospitality.
D.Anselmi is supported by the Chinese Academy of Sciences visiting
professorship for senior international scientists, grant No. 2010T2J01.

\vskip 20truept \noindent {\Large \textbf{Appendix A: Polar decomposition
and diagonalization of matrices}}

\vskip 10truept

\renewcommand{\theequation}{A.\arabic{equation}} \setcounter{equation}{0}

In this appendix we review some definitions and results about the polar
decomposition of matrices and their diagonalization. We present them in ways
that are useful for the arguments of our paper.

\medskip \textbf{Definition} \textit{Let $g\in U(n)$ be a unitary $n\times n$
matrix and $h$ a diagonal unitary matrix, namely an element of the subgroup $%
U(1)^{n}\subset U(n)$. Consider the set of left cosets of $U(1)^{n}$ in $%
U(n) $, namely the equivalence classes under the equivalence relation: $%
g\sim g^{\prime }$ if and only if $g^{-1}g^{\prime }=h\in U(1)^{n}$. This
set is denoted with $\tilde{U}_{\ell }(n)$. Its real dimension is $n(n-1)$.}

\medskip \textbf{Theorem 2} \textit{Let $H$ be a Hermitian $n\times n$
matrix. There exists a diagonal matrix $D= $diag$(d_{1},\cdots ,d_{n})$ with 
$d_{1}\geqslant d_{2}\geqslant \cdots \geqslant d_{n}$ and a unitary matrix $%
\tilde{U}$ belonging to $\tilde{U}_{\ell }(n)$, such that} 
\begin{equation}
H=\tilde{U}D\tilde{U}^{\dagger }.  \label{diagonal}
\end{equation}

The diagonal unitary matrices of $U(1)^{n}$ commute with $D$, so they do not
contribute to (\ref{diagonal}). The diagonalization (\ref{diagonal}) is
unique if $H$ does not have degenerate eigenvalues. We can prove this
statement checking that the dimensions match: the set of Hermitian matrices
has real dimension $n^{2}$, which is equal to the sum of the dimension of $%
\tilde{U}_{\ell }(n)$, which is $n^{2}-n$, plus the dimension of the set of
diagonal matrices $D$, which is $n$.

Now we consider the polar decomposition of matrices, which we present in a
form that is again generically unique.

\medskip \textbf{Theorem 3} \textit{Let $S$ be any invertible complex $%
n\times n$ matrix. There exists a non-negative diagonal matrix $D=$diag$%
(d_{1},\cdots ,d_{n})$ with $d_{1}\geqslant d_{2}\geqslant \cdots \geqslant
d_{n}>0$, and matrices $U_{L}$, $\tilde{U}_{R}$ belonging to $U(n)$ and $%
\tilde{U}_{\ell }(n)$, respectively, such that}\footnote{%
The reason why ``R'' stands to the left and ``L'' stands to the right in
formula (\ref{su}) is that in this way $U_{L}$ is attached to left-handed
quarks and $\tilde{U}_{R}$ is attached to right-handed quarks, according to (%
\ref{LY}).} 
\begin{equation}
S=\tilde{U}_{R}DU_{L}.  \label{su}
\end{equation}

\textit{Proof}. Since $S$ is invertible, we can write 
\begin{equation}
S=SS^{\dagger }(S^{\dagger })^{-1}.  \label{b}
\end{equation}
Now, $SS^{\dagger }$ is Hermitian, so it can diagonalized with a unitary
matrix $\tilde{U}_{R}\in \tilde{U}_{\ell }(n)$. Since $SS^{\dagger }$ is
also positive-definite, we call its diagonal form $D^{2}$ and define $D$ as
the positive square root of $D^{2}$. We have 
\begin{equation}
SS^{\dagger }=\tilde{U}_{R}D^{2}\tilde{U}_{R}^{\dagger }.  \label{c}
\end{equation}
Inserting (\ref{c}) in (\ref{b}) we get (\ref{su}) with 
\[
U_{L}=D\tilde{U}_{R}^{\dagger }(S^{\dagger })^{-1}. 
\]
This matrix is unitary. Indeed, 
\[
U_{L}^{\dagger }U_{L}=S^{-1}\tilde{U}_{R}D^{2}\tilde{U}_{R}^{\dagger
}(S^{\dagger })^{-1}=1. 
\]

Again, the dimensions match, because $S$, $\tilde{U}_{R}$, $D$ and $U_{L}$
contain $2n^{2}$, $n^{2}-n$, $n$ and $n^{2}$ real parameters, respectively.
Thus, if the eigenvalues of $SS^{\dagger }$ are non-degenerate the
decomposition is unique.

Finally, consider the Hermitian matrix 
\[
N=\left( 
\begin{tabular}{cc}
$0$ & $S^{\dagger }$ \\ 
$S$ & $0$%
\end{tabular}
\right) . 
\]
Using (\ref{su}), we can diagonalize it with the unitary matrix, 
\[
U=\frac{1}{\sqrt{2}}\left( 
\begin{tabular}{cc}
$U_{L}^{\dagger }$ & $U_{L}^{\dagger }$ \\ 
$\tilde{U}_{R}$ & $-\tilde{U}_{R}$%
\end{tabular}
\right) . 
\]
The eigenvalues of $N$ come in pairs of opposite signs, and coincide with
the diagonal entries of $D$ and their opposites: 
\begin{equation}
N=U\left( 
\begin{tabular}{cc}
$D$ & $0$ \\ 
$0$ & $-D$%
\end{tabular}
\right) U^{\dagger }.  \label{diago}
\end{equation}
\vskip 20truept \noindent {\Large \textbf{Appendix B: Mathematical
definitions}}

\vskip 10truept

\renewcommand{\theequation}{B.\arabic{equation}} \setcounter{equation}{0}

Here we collect some mathematical definitions used in the paper. The
calculation of our one-loop diagrams gives the functions 
\begin{equation}
f_{i_{1}\cdots i_{n}}=\frac{(n-1)!}{(4\pi )^{2}}\int_{0}^{1}\mathrm{d}%
x_{1}\int_{0}^{1-x_{1}}\mathrm{d}x_{2}\cdots
\int_{0}^{1-\sum_{k=1}^{n-2}x_{k}}\mathrm{d}x_{n-1}\left( \ln \frac{\Lambda
_{L}^{2}}{M_{n,x}^{2}}+c_{n}\right) ,  \label{ff}
\end{equation}
where $i_{1}\cdots i_{n}$ can have the values $t$ and $b$, 
\[
M_{n,x}^{2}=\sum_{k=1}^{n-1}m_{i_{k}}^{2}x_{k}+m_{i_{n}}^{2}\left(
1-\sum_{k=1}^{n-1}x_{k}\right) 
\]
and $c_{n}$ are constants. The first constants $c_{n}$ have approximate
numerical values 
\[
c_{2}=-2.11371,\qquad c_{3}=-2.61371,\qquad c_{4}=-2.94704. 
\]

The diagrams are calculated as follows. Using the gap equation, the momentum
integrals are convergent for $\Lambda _{L}<\infty $ and logarithmically
divergent when $\Lambda _{L}$ is sent to infinity. They can be viewed as
regularized by the Lorentz violation. A direct evaluation of Lorentz
violating integrals is very difficult. However, renormalization theory
ensures that everything but finite numerical constants (the constants $c_{n}$
) can be unambiguously calculated with any regularization method. We used an
ordinary cut-off. Later, we evaluated the constants $c_{n}$ taking equal
masses in the Lorentz violating integrals.

With $\Lambda _{L}=10^{14}$GeV and the known values of $m_{t,b}$ we see that
the constants $c_{n}$ are numerically not important for the analysis of our
paper.

Clearly, $f_{i_{1}\cdots i_{n}}$ is completely symmetric. Using $m_{b}\ll
m_{t}\ll \Lambda _{L}$, we have 
\[
f_{i_{1}\cdots i_{n}}\sim \frac{1}{(4\pi )^{2}}\ln \frac{\Lambda _{L}^{2}}{%
m_{t}^{2}}, 
\]
any time at least one index is $t$. Instead, 
\[
f_{b\cdots b}\sim \frac{1}{(4\pi )^{2}}\ln \frac{\Lambda _{L}^{2}}{m_{b}^{2}}%
. 
\]

Note the change of notation with respect to \cite{noH}, because we have
expanded all functions contained in the low-energy effective action in
powers of the momentum

\end{document}